\documentclass[final, 5p, sort&compress, times, twocolumn]{elsarticle}

\usepackage{multirow}
\usepackage{hyperref}
\usepackage{booktabs}
\usepackage{framed} 
\usepackage{multicol} 
\usepackage{nomencl} 
\usepackage{afterpage}
\usepackage[numbers]{natbib}
\usepackage{tabularx}

\usepackage{makecell}
\usepackage{cellspace}
\usepackage[justification=justified,font=footnotesize,skip=0pt]{caption}
\usepackage[binary-units=true]{siunitx}
\DeclareSIUnit \VAr {VAr} 
\DeclareSIUnit \VA {VA} 
\DeclareSIUnit \rad {Radians} 

\usepackage{mathtools}
\usepackage{algorithm, algorithmic}
\usepackage[font=footnotesize]{caption,subfig}
\usepackage{multirow}

\usepackage{enumitem}
\usepackage{xcolor}
\usepackage{lipsum}
\usepackage{comment}
\allowdisplaybreaks

\usepackage{stackengine,scalerel}

\newcommand\obullet[1]{\ThisStyle{\ensurestackMath{%
  \stackon[1pt]{\SavedStyle#1}{\SavedStyle\kern.6\LMpt\bullet}}}}
\newcommand\ocirc[1]{\ThisStyle{\ensurestackMath{%
  \stackon[1pt]{\SavedStyle#1}{\SavedStyle\kern.6\LMpt\circ}}}}

\usepackage{cleveref}
\crefname{equation}{equation}{equations}
\Crefname{equation}{Equation}{Equations}
\crefrangelabelformat{equation}{(#3#1#4--#5#2#6)}

\crefmultiformat{equation}{equations (#2#1#3}{, #2#1#3)}{#2#1#3}{#2#1#3}
\Crefmultiformat{equation}{Equations (#2#1#3}{, #2#1#3)}{#2#1#3}{#2#1#3}

\usepackage{amsmath}
\usepackage{mathdots}
\usepackage{yhmath}
\usepackage{cancel}
\usepackage{color}
\usepackage{tikz}
\usepackage{circuitikz}
\usetikzlibrary{positioning,chains,fit,shapes,calc,arrows,shadows}
\usepackage{array}
\usepackage{amssymb}
\usepackage{gensymb}
\usepackage{tabularx}
\usepackage{booktabs}
\usetikzlibrary{fadings}
\usetikzlibrary{patterns}
\usetikzlibrary{shadows.blur}

\def\tsc#1{\csdef{#1}{\textsc{\lowercase{#1}}\xspace}}
\tsc{WGM}
\tsc{QE}
\tsc{EP}
\tsc{PMS}
\tsc{BEC}
\tsc{DE}


\begin{document}

\begin{frontmatter}



\title{Can phase change materials in building insulation improve self-consumption of residential rooftop solar? An Australian case study}     

\author[1]{Zahra Rahimpour\corref{cor1}}
\address[1]{School of Electrical and Information Engineering, The University of Sydney, Sydney, Australia}
\cortext[cor1]{Corresponding author}
\ead{zahra.rahimpour@sydney.edu.au}

\author[1]{Gregor Verbi\v{c}}
\ead{gregor.verbic@sydney.edu.au}

\author[2]{Archie C. Chapman}
\ead{archie.chapman@uq.edu.au}

\address[2]{School of Information Technology and Electrical Engineering, The University of Queensland, Brisbane, Australia}

\begin{abstract}
This work investigates the extent to which \textit{phase change material} (PCM) in the building's envelope can be used as an alternative to battery storage systems to increase self-consumption of rooftop solar photovoltaic (PV) generation. In particular, we explore the electricity cost-savings and increase in PV self-consumption that can be achieved by using PCMs 
and the operation of the heating, ventilation, and air conditioning (HVAC) system optimised by a \textit{home energy management system} (HEMS). In more detail, we consider a HEMS with an HVAC system, rooftop PV, and a PCM layer integrated into the building envelope. The objective of the HEMS optimisation is to minimise electricity costs while maximising PV self-consumption and maintaining the indoor building temperature in a preferred comfort range. Solving this problem is challenging due to PCM's nonlinear characteristics, and using methods that can deal with the resulting non-convexity of the optimisation problem, like \textit{dynamic programming} is computationally expensive. Therefore, we use \textit{multi-timescale approximate dynamic programming} (MADP) that we developed in our earlier work to explore a number of Australian PCM scenarios. 
Specifically, we analyse a large number of residential buildings across five Australian capital cities. We find that using PCM can reduce annual electricity costs by between \SI{10.6}{\%} in Brisbane and \SI{19}{\%} in Adelaide. 
However, somewhat surprisingly, using PCM reduces PV self-consumption by between \SI{1.5}{\%} in Brisbane and \SI{2.7}{\%} in Perth.

\end{abstract}

\begin{keyword}
Building thermal inertia \sep 
demand response  \sep 
electricity cost-saving \sep 
home energy management \sep 
phase change material \sep 
PV self-consumption.
\end{keyword}

\end{frontmatter}

\section{Introduction}
\label{sec:Introduction}
Reducing costs of rooftop solar photovoltaic (PV) and government policies that support PV uptake have resulted in a dramatic increase in the deployment of rooftop solar PV across Australia over the past ten years. 
As a result, there are now over two million residential rooftop solar systems with a total capacity exceeding \SI{8}{\giga\watt} \cite{Graham2019}. 
On the other hand, reduced feed-tariffs (the financial remuneration for exporting excess solar generation to the grid) are driving householders' interest in increase PV self-consumption, with battery storage being the most popular choice. 
However, battery storage is still expensive; hence there is an interest in other loads that can serve as a \emph{solar sponge}. 
One possible candidate is space heating and cooling, which constitutes between \SI{20}{\%}-\SI{40}{\%} (or even up to \SI{50}{\%} in certain countries) of total energy consumption \cite{perez2008review}.

A building's thermal inertia---the ability of a building envelope to store or release thermal energy---has been demonstrated to be an effective means to reduce or shift the heating and cooling demand \cite{verbeke2018thermal}. 
However, in some countries such as Australia, buildings have a lightweight construction and hence low thermal inertia \cite{gregory2008effect,rahimpour2017using}, which makes them ineffective for load shifting. 
In these settings, integrating \textit{phase change material} (PCM) into a building's envelope improves its thermal inertia significantly. PCMs have a higher volumetric heat capacity compared to materials with high thermal inertia, such as bricks. When the temperature changes, PCMs store and release heat by undergoing a phase transition from solid to liquid or vice versa, which is akin to the energy charging of electrochemical batteries. 
Storing or releasing latent heat during phase transition provides the building with a sufficient thermal mass to smooth indoor temperature fluctuations, which can reduce or shift the building's heating and cooling demand.

In this paper, we consider a plant-based PCM\footnote{https://phasechange.com/biopcm/}, which is more environmentally friendly and more easily disposed compared to electrochemical batteries. 
Moreover, it has a long lifetime (almost 80 years) compared to batteries (about 15 years). Another benefit is that PCMs are easy to install, making them suitable for retrofitting buildings.

In the next section, we briefly review the relevant literature on the use of PCM for the management and control of the  thermal performance of residential buildings.

\subsection{Background} \label{Sec:Intro_Background}
The ability of PCMs to reduce peak heating and cooling demand and improve indoor thermal comfort has attracted a lot of attention in research literature \cite{song2018review}. However, the literature shows that achieving an effective PCM performance is difficult, and it is affected by several factors, including the climate, building parameters, and the properties of the PCM employed \cite{al2020incorporation}. For instance, using PCM is not beneficial in hot and humid climates with insufficient diurnal temperature variation to solidify and melt the PCM \cite{al2020peak}. Therefore, passive application\footnote{Passive application refers to the use of PCM where the phase change occurs without the aid of mechanical devices such as a HVAC system.} of PCM cannot always maintain the indoor temperature in the desired comfort range \cite{al2020peak}.
On the other hand, active techniques can efficiently control the buildings’ thermal performance, especially in locations where passive PCM application has a limited potential. For example, the authors in \cite{bai2020analytical} indicated that overheating cannot be avoided by simply using PCMs; instead, ventilation is needed to unleash the full PCM potential.

Although active use of a PCM helps to exploit its storage potential effectively, its efficiency depends on control and implementation strategies \cite{barzin2015application}. In more detail, optimising the PCM performance needs to be defined as an optimal control problem. Moreover, it is critical how we define the optimal PCM performance. For example, the optimal PCM melting point that achieves the highest resiliency of the building to extreme weather conditions yields only around \SI{60}{\%} of energy-saving compared to optimising the PCM's performance for energy-saving \cite{baniassadi2019effectiveness}. 

The literature on optimal control of buildings with PCMs is scarce. Most of the existing research uses simulation-based optimisation \cite{saffari2017simulation}, with the problem variables typically including PCM properties, building's envelope properties, or the HVAC system operating conditions. The optimal solution is found iteratively by optimising the objective function with different values of the input variables so that in each iteration the solution moves closer to the optimum. Because the set of possible input variables is limited, the optimal solution is not optimal globally. The other drawback is that simultaneously considering all the variables that affect the objective function is either infeasible or time-consuming, or requires a lot of of trial and error. For example, the authors in \cite{barzin2015application} used price-based control to switch on/off the underfloor heating system in two identical test huts. They demonstrated that using PCM results in both peak load shifting and electricity saving.  However, they only considered the electricity price as a signal to operate the heating system without considering other factors that contribute to optimal control of the heating system, such as the optimal performance of the PCM. 

The other widely-used optimisation method in PCM buildings is inverse problem-based simulation.  In this method, reverse engineering is used to find an optimal solution. The desired results are considered an optimisation objective, and variables of the problem are adjusted to achieve the said objective. For instance, the authors in \cite{ cheng2013new} applied inverse problem-based optimisation to determine the optimal thermophysical properties of a PCM-concrete brick. The distribution of the specific heat of the envelope structure with temperature is used as the optimisation variable. They continuously adjust the distribution of specific heat with temperature using the sequential quadratic programming (SQP) method until the optimisation objective is achieved. In addition, they also compare this solution with particle swarm optimisation and a genetic algorithm. They concluded that SQP gives a better solution than two other methods. 

In addition to the lack of efficient, powerful optimisation methods for PCM applications, the second literature gap is the co-optimisation of PCMs with other distributed energy resources such as rooftop PV, particularly in the context of maximising PV self-consumption. To the best of our knowledge, \cite{sun2022optimized} is the only paper investigating the use of PCMs in building insulation in conjunction with other distributed energy resources. Specifically, the authors in \cite{sun2022optimized} investigated energy-saving and electricity demand shifting that can be achieved using insulation boards with PCM in a lightweight building equipped with a PV system and a battery storage system. In more detail, they simulated a $\SI{2.5}{\metre}\times\SI{2.5}{\metre}\times\SI{2.5}{\metre}$ cubic chamber equipped with an air conditioning system. They simulated the thermal behaviour of the building in EnergyPlus software and validated the simulation results with an experiment. Considering the building with a PV system without PCM as a baseline case, they observed a \SI{47}{\%} reduction in peak cooling load and an hour shift in the cooling demand in the summer.
The battery charging and discharging is determined by a simple heuristic to maximise PV self consumption. In more detail, the HVAC system is supplied by PV first, and when the PV generations exceed the HVAC consumption, the excess energy is stored in the battery. When the PV generation is insufficient, the energy is taken from the battery, and when the battery is empty, the power is taken from the grid. The authors tested the performance of the test chamber with different battery PV system sizes. The main drawback of \cite{sun2022optimized} is the lack of a principled optimisation to achieve the optimal HVAC performance in a more realistic setting with and underlying electricity consumption and time-varying electricity prices. Also, \cite{sun2022optimized}  only considered a single summer day, which leaves the question of the PCM performance throughout the year.

The main challenge in optimising the PCM's performance is the highly nonlinear specific heat capacity characteristics \eqref{straincomponent}, which throws up a couple of challenges when using Newton-based methods like SQP. First, \eqref{straincomponent} is nonsmooth at the melting point, requiring computationally expensive evaluation of derivatives (either through black-box simulations and/or via finite-differencing) \cite{Byrd2006}. Second, the resulting optimisation problem is highly nonconvex, so choosing a good starting point to prevent the algorithm from getting trapped in a local optimum is not trivial \cite{Boggs2000}. Given this, we use dynamic programming, which can handle the nonlinearity introduced by the PCM. However, dynamic programming suffers from the \textit{curse of dimensionality} \cite{Bellman2015}. Therefore, we use our previously-developed \textit{multi-time scale approximate dynamic programming} to reduce the computational burden of dynamic programming while maintaining the quality of the solution \cite{rahimpour2021comput}. 

Against this background, this work illustrates how PCMs can be used as an alternative to battery storage to increase PV self-consumption and reduce electricity costs in the context of home energy management.
We use five Australian capital cities with different climatic conditions as a case study.
We cast the problem as an optimisation problem, which is solved by an automated \textit{home energy management system} (HEMS). 
The HEMS consists of an HVAC system, rooftop PV, and a PCM layer integrated into the building envelope. 
A HVAC schedule is determined by the HEMS solving an optimisation problem, with the objective of minimising the home's electricity cost, while maximising PV self-consumption and maintaining the building's thermal comfort. 

\subsection{Contributions of the paper}
The paper's main contribution is a techno-economic analysis of the viability of PCMs as an alternative to battery storage to increase PV self-consumption and reduce electricity costs. 
To our knowledge, this work is the first attempt to: 
\begin{enumerate}
    \item Illustrate the potential economic benefits of PCMs for a large number of residential buildings (210 dwellings in total). The benefits of using PCM are evaluated and quantified by calculating (i) electricity cost-saving (ii) increase in PV self-consumption.  
    \item Exploit the optimal performance of PCMs by defining the problem as an optimisation problem and solved using a powerful and computationally efficient method that can handle the nonlinearity of the optimisation problem. 
    \item Comprehensively cover different climatic conditions and the variability of end-user electricity demand and PV generation. To that end, we use historical data (the calendar year of 2019) of weather, PV generation and residential demand.
\end{enumerate}

\subsection{Outline of the paper} 
   
This paper progresses as follows: in Section \ref{sec:Thermal RC lumped model of PCM-buildings}, a thermal model of a HEMS-PCM
is briefly described. In Section \ref{sec:Home energy management in HVAC PV-PCM buildings}, the optimisation problem of the HEMS-PCM is formulated, and the proposed MADP algorithm is used to solve it. 
Section~\ref{sec:Case studies} describes the case studies used in simulations. Section~\ref{sec:Results and discussion} presents and discusses the simulation results. Section~\ref{sec:Conclusion} concludes the paper and outlines future directions of this work.

\section{Thermal model of PCM-buildings}       
\label{sec:Thermal RC lumped model of PCM-buildings}       
To formulate the optimisation problem of HEMS-PCM, we require a dynamic thermal model of the building. 
We consider a single zone building with dimensions of $\SI{8}{\metre}\times\SI{6}{\metre}\times\SI{2.7}{\metre}$ and a total floor area of $\SI{48}{\metre\squared}$. The materials and configuration are chosen such that the building is representative of lightweight construction in Australia \cite{gregory2008effect}. 
The building elements
are constructed of an outer layer of rendered fibro-cement, followed by a timber stud wall containing insulation batts, with plasterboard on the interior \cite{gregory2008effect}. The parameters are given in Table~\ref{table1}, and the fenestration is detailed in Table~\ref{table3}. 

To improve the thermal inertia of the building, we add a \SI{0.03}{\metre} PCM layer underneath the plasterboard layer in all elements of the building (roof, walls, and floor). The properties of the type of PCM used in this work are detailed in Table~\ref{table2}. 

The building is equipped with a \SI{4}{\kilo\watt} HVAC system that, based on optimal scheduling, operates either with power drawn from the grid or electricity generated by the rooftop solar PV system. 
The coefficient of performance (COP) of the HVAC system is assumed 4.5\footnote{The COP is a function of the difference between indoor and outdoor temperatures. Although Australia does experience heat waves as well as occasional cold snaps, the temperature variation between winter and summer is modest, hence we assumed a constant COP.}. This means for each \si{\kilo\watt\hour} of electrical energy used by the HVAC system, \SI{4.5}{\kilo\watt\hour} worth of thermal energy can be removed from or added to the building. To further simplify the simulations, we neglect the heat gain from the occupants and other internal heat gains. Moreover, we disregard the solar gains and the effect of wind. 

\begin{table}
    \footnotesize
	\centering
	\caption{Building elements composition and its material properties \cite{handbook2009american,gregory2008effect}.}
	\label{table1}
	\begin{tabular}{lcccc} 
		\toprule
		\multirow{2}{*}[0em]{\textbf{Element}} & $d$ & $\lambda$ &  $\rho$ &  $c$  \\
		& (m) & (\si{\watt / \meter \kelvin}) & (\si{\kilogram / \meter \cubed}) & (\si{\joule / \kilogram. \kelvin}) \\
		\midrule
		Rendered  & \multirow{2}{*}[0em]{0.005} & \multirow{2}{*}[0em]{0.25} & \multirow{2}{*}[0em]{1150} & \multirow{2}{*}[0em]{840} \\
		fibro-cement & & & & \\
		Timber studwall & \multirow{2}{*}[0em]{0.09} & \multirow{2}{*}[0em]{0.15} & \multirow{2}{*}[0em]{650} & \multirow{2}{*}[0em]{1200} \\
		with insulation batts & & & & \\ 
		Plaster board & 0.01 & 0.25 & 950 & 840 \\
		\bottomrule
	\end{tabular}
\end{table}

\begin{table}[t]
	\footnotesize
	\centering
	\caption{Fenestration details \cite{evola2011simulation}.}
	\label{table3}
	\begin{tabular}{llcc} 
		\toprule
		\begin{tabular}[c]{@{}l@{}}\textbf{Element}\\\end{tabular} & \textbf{Description} & \begin{tabular}[c]{@{}c@{}}$U$\\ (\si{\watt / \meter \squared \kelvin}) \end{tabular} & \begin{tabular}[c]{@{}c@{}} Area\\($\SI{}{\metre\squared}$)\end{tabular} \\
		\midrule
		\begin{tabular}[c]{@{}l@{}}Windows\\\end{tabular} & \begin{tabular}[c]{@{}l@{}}Single glazing with Aluminium frame\end{tabular} & 7.01 & \begin{tabular}[c]{@{}l@{}}7.8\end{tabular} \\
		Door & \begin{tabular}[c]{@{}l@{}}Wooden slab with wooden frame\end{tabular} & 2.61 & 2.1 \\
		\bottomrule
	\end{tabular}
\end{table}

\begin{table}[t]
	\footnotesize
	\centering
	\caption{Honeycomb PCM properties \cite{evola2011simulation}.}
	\label{table2}
	\begin{tabular}{@{}lcccc@{}} 
		\toprule
		\begin{tabular}[c]{@{}l@{}}\textbf{PCM type}\\\end{tabular} & $d$ (\si{\meter}) & \begin{tabular}[c]{@{}l@{}}\textbf{$\lambda$} \textbf{(\si{\watt / \meter \kelvin})}\end{tabular} & \begin{tabular}[c]{@{}l@{}}\textbf{$\rho$} \textbf{( \si{\kilogram / \meter \cubed })}\end{tabular} & $c$ (\si{\joule / \kilogram. \kelvin}) \\
		\midrule
		Honeycomb PCM & 0.03 & 2.8 & 545 & varies \\
		\bottomrule
	\end{tabular}
\end{table}

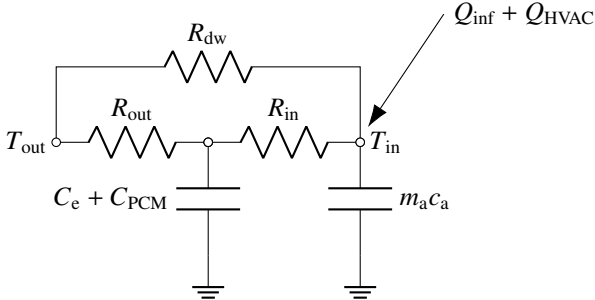
\begin{figure}[t!]
	\centering
	\begin{circuitikz}[american voltages]
		\draw (0,0) to [short] (0,1.0) to [R=$R_\text{dw}$] (4,1.0) to [short] (4,0);
		\draw (0,0) to[R=$R_\text{out}$,o-o] (2,0) to [R=$R_\text{in}$,o-o] (4,0);
		\draw[{Latex[length=3mm, width=2mm]}-] (4.1,0.2) -- (5.1,1.7) node[right] {$Q_\text{inf}+Q_\text{HVAC}$};
		\draw (2,-1.5) node[ground]{} to[C=$C_\text{e}+C_\text{PCM}$,-o] (2,0);
		\draw (4,0) to[C=$m_\text{a}$$c_\text{a}$,o-] (4,-1.5) node[ground]{}
		(0,0) node [left]{$T_\text{out}$}
		(4,0) node [right]{$T_\text{in}$};
	\end{circuitikz}
	\caption{ 2RC lumped model of PCM-building }
	\label{2RCpcm}
\end{figure}

We use the lumped RC thermal model from our earlier work \cite{rahimpour2018APVI}. The model is shown in Fig.~\ref{2RCpcm}. Using a lumped RC modelling approach is typical in whole-building energy simulation \cite{underwood2008modelling}. In this approach, each building element is considered a single dimensionless lump by assuming a uniform temperature across the building's element \cite{underwood2008modelling}. 
Due to the close resemblance of the thermal energy balance to Ohm's law, electric resistance and capacitance can be considered analogous to thermal resistance and capacitance of the building. We built the model in Matlab and validated its performance by benchmarking it against the widely used EnergyPlus software. Note that placing the PCM underneath the plasterboard allows us to add together the thermal capacitances of the building envelope and the PCM so that the thermal mass of the building envelope in represented by a single thermal capacitance.
In this arrangement, the thermal resistance between the PCM layer and plasterboard is very small and can be neglected. Therefore, the PCM's surface temperature and the building's indoor temperature are approximately the same, so monitoring the indoor temperature is sufficient for analysing the thermal performance of the PCM. 

The differential equations governing the thermal behaviour of the building are given by:
\begin{align}
    \dot{T_\mathrm{e}} &= \frac{1}{C_\mathrm{e}+C_\mathrm{PCM}}\left(\frac{T_\mathrm{in} - T_\mathrm{e}}{R_\mathrm{in}}+\frac{T_\mathrm{out} - T_\mathrm{e}}{R_\mathrm{out}}\right), \label{eqn7} \\
    \dot{T}_\mathrm{in} &= \frac{1}{m_\mathrm{a}c_\mathrm{a}}\left(\frac{T_\mathrm{out} - T_\mathrm{in}}{R_\mathrm{dw}}+\frac{T_\mathrm{e} - T_\mathrm{in}}{R_\mathrm{in}}+\dot{Q}_\mathrm{HVAC} + \dot{Q}_\mathrm{inf}\right), \label{eqn8}
\end{align}
where $R_\mathrm{in}$ and ${R}_\mathrm{out}$ represent thermal resistance of inner and outer layers of the element. The total thermal resistance of the building’s envelope (excluding doors and windows) is split into inner and outer thermal resistance, and the accessibility factor introduced in \cite{masy2008definition} is used to calculate the inside and outside thermal resistances. The fenestration (doors and windows) is represented by a pure thermal resistance (${R}_\mathrm{dw}$) due to its negligible thermal mass.

The total capacitance of the building and the PCM are represented by $C_\mathrm{e}$, and $C_\mathrm{PCM}$ respectively.  Moreover, the thermal capacity of the indoor air is accounted for by the term $m_\mathrm{a}c_\mathrm{a}$.  The outdoor temperature, indoor temperature and the surface temperature of PCM are captured by $T_\mathrm{out}$, $T_\mathrm{in}$ and $T_\mathrm{e}$, respectively. The energy from the HVAC system and infiltration energy that enters the living space are incorporated by the terms $Q_\mathrm{HVAC}$ and $Q_\mathrm{inf}$, respectively. 

\begin{figure}[t]
	\centering
	\centerline{\includegraphics[width=80mm,keepaspectratio]{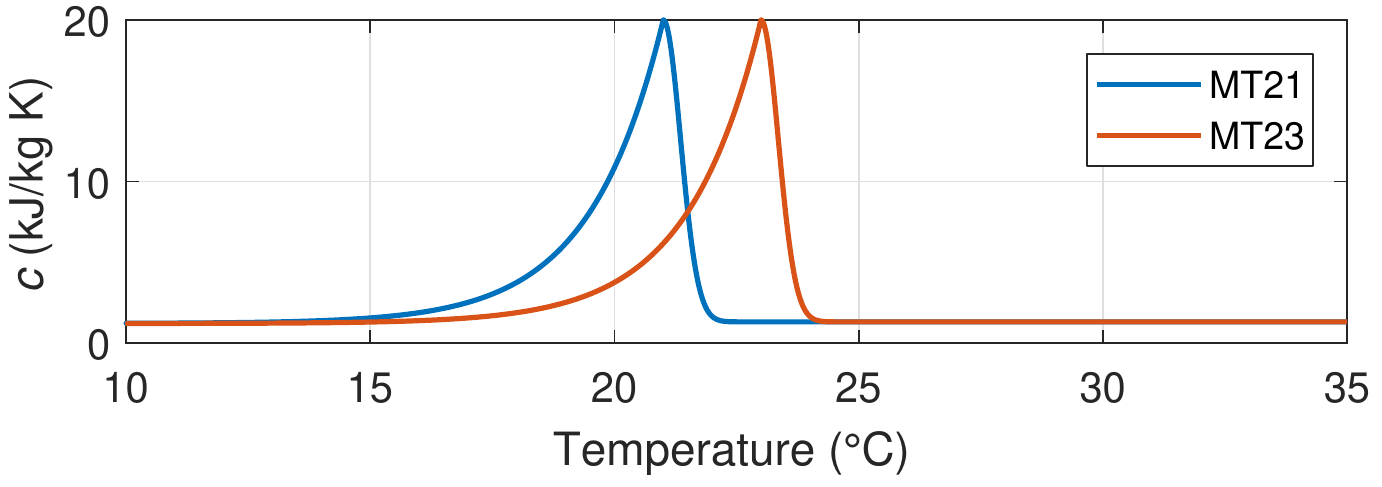}}
	\caption{Specific heat capacity characteristics of PCMs with melting points of \SI{21}{\degreeCelsius} (MT21, blue), and \SI{23}{\degreeCelsius} (MT23, red).}
	\label{cpcm}
\end{figure}

We consider two different PCM melting points (the temperature at which the heat capacity is the highest): \SI{21}{\degreeCelsius} (MT21) and \SI{23}{\degreeCelsius}  (MT23). The specific heat capacity as a function of temperature for the two PCMs, shown in Fig.~\ref{cpcm}, can be approximated by the following function:
\begin{subequations} \label{eqn6}
	\begin{align}
		\begin{split} \label{eqn6a}
			{c_\mathrm{pcm}}&= 1200 + 18800e^{-\left(\frac{{T_\mathrm{p}} - {T}}{1.5}\right)}  \quad \mathrm{if} \quad  {T} < {T_\mathrm{p}},   \\
		\end{split}\\
		\begin{split} \label{eqn6b}
			{c_\mathrm{pcm}}&= 1300 + 18700e^{-4\left({T_\mathrm{p}} - {T}\right)^{2}}  \quad \mathrm{if} \quad  {T} \geq {T_\mathrm{p}},  \\
		\end{split}
	\end{align}
	\label{straincomponent}
\end{subequations}
$\!\!$where ${T_\mathrm{p}}$ is the melting point of the PCM. Observe that the phase change for MT21 occurs over the range \SI{15.4}{\degreeCelsius} to \SI{21.9}{\degreeCelsius}, while for MT23 the range is between \SI{17.4}{\degreeCelsius} to \SI{23.9}{\degreeCelsius}; this can be thought of as the \emph{operating range} of the PCM where its ability to store and release heat is the highest.

\subsection{PCM as an energy storage medium}
The PCM acts as an energy storage medium that absorbs and stores heat, which enables preheating and precooling of the building. To quantify the amount of heat the PCM can store we need to introduce enthalpy. The enthalpy of a system is the sum of the system’s internal energy and the product of its pressure and volume. Since the pressure and volume are constant, the variation in enthalpy equals the variation in the internal energy of the PCM. By calculating the change in enthalpy as the temperature of the PCM changes from $T_1$ to $T_2$, we can assess the amount of energy stored or released from the PCM as follows:
\begin{align}\label{eqn15}
	\Delta H=\int_{{T}_{1}}^{{T_2}} {m_\mathrm{pcm}}{c}_\mathrm{pcm}\left(T\right) dT \quad \textrm{(J)},
\end{align}
where ${c_\mathrm{pcm}}$ is the specific heat capacity in \si{\joule\per\kilo\gram\per\kelvin} illustrated in Fig.~\ref{cpcm}, and ${m_\mathrm{PCM}}$ is the total mass of the PCM in kg used in the building envelope (\SI{2806}{\kilo\gram} in our case). 

\begin{figure}[t]
	\centering
	\centerline{\includegraphics[width=80mm,keepaspectratio]{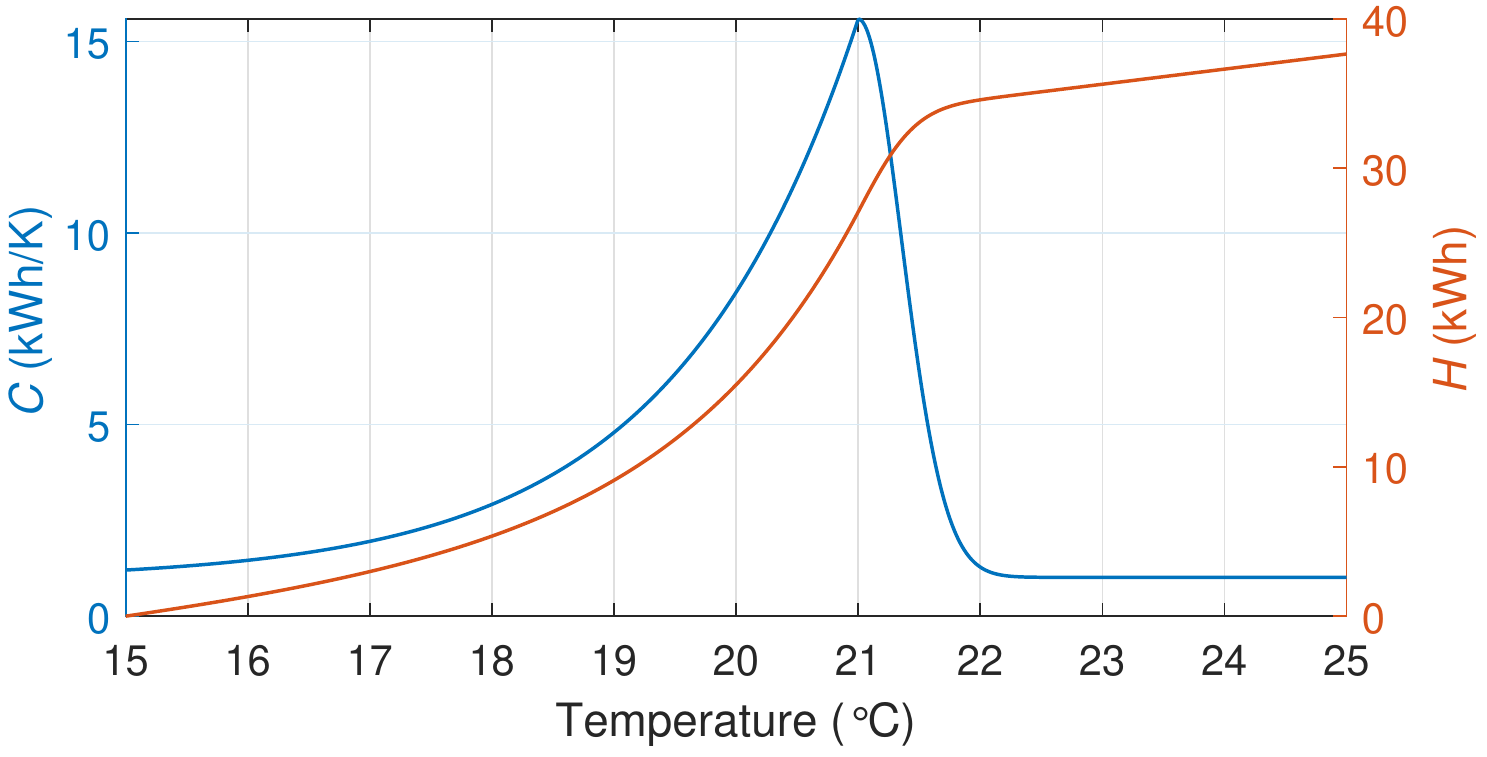}}
	\caption{Total heat capacity $C$ (left axis) and enthalpy $H$ (right axis) of a PCM with a melting point of \SI{21}{\degreeCelsius}. (Note that these are \emph{total} not \emph{specific} values.)}
	\label{fig_PCM_enthalpy}
\end{figure}

The enthalpy as a function of temperature along with the specific heat capacity for the melting point of \SI{21}{\degreeCelsius} is illustrated in Fig.~\ref{fig_PCM_enthalpy}. For ease of comparison, we have converted the units to more familiar units of kilowatt hours (kWh). When the PCM temperature changes from \SI{15}{\degreeCelsius} to \SI{25}{\degreeCelsius}, the PCM stores almost \SI{40}{\kilo\watt\hour} worth of thermal energy. Considering only the operating range of the PCM between \SI{20}{\degreeCelsius} to \SI{24}{\degreeCelsius} (occupant comfort temperature range), the storage capacity of the PCM is about \SI{21}{\kilo\watt\hour}. However, to be able to directly compare that to a storage capacity of an electrochemical battery, we need to consider the COP of the air conditioner to convert the thermal energy into equivalent electrical energy of the HVAC system. Thus, assuming a COP of 4.5, the PCM can store an equivalent of \SI{5}{\kilo\watt\hour} worth of electrical energy of the HVAC system (assuming no leakage of thermal energy). That is, of course, a crude approximation, which nevertheless gives a sense of the PCM's energy storage capacity.

\begin{figure}[t]
	\centering
	\centerline{\includegraphics[width=90mm,keepaspectratio]{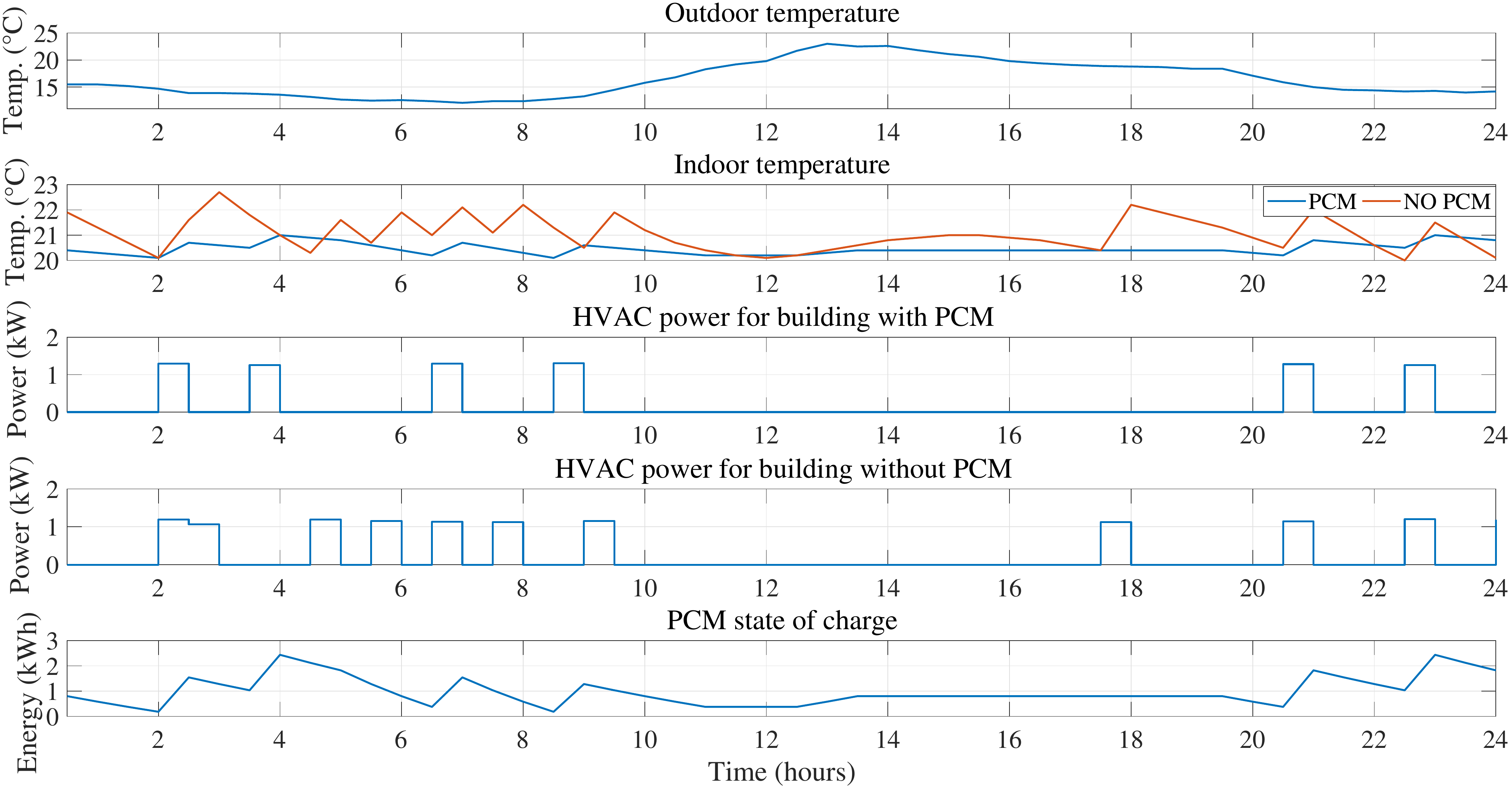}}%
			\caption{Thermal performance of a building for a typical winter day in Sydney. From top to bottom: outdoor temperature, indoor temperature of a building with PCM vs a building without PCM, HVAC power for HEMS with PCM, HVAC power for HEMS without PCM, and PCM state of charge.}
	\label{fig:SOC}
\end{figure}

To further illustrate how a PCM acts as an energy storage medium, we show the thermal performance of a building in Sydney on a typical winter day in Fig.~\ref{fig:SOC}. The figure compares the indoor temperature and the HVAC system operation of a building with a PCM versus the same building without PCM. The PCM \textit{state of charge} (SOC) on the bottom plot shows the amount of energy stored in the PCM. The energy is corrected for the COP, so the SOC varies between \SI{0.2}{\kilo\watt\hour} and \SI{2.5}{\kilo\watt\hour} for the indoor temperature between \SI{20.1}{\degreeCelsius} and \SI{21}{\degreeCelsius} experienced on the day (assuming SOC is zero at \SI{20}{\degreeCelsius}). Observe how the operation of the HVAC systems increases the SOC of the PCM without a noticeable change in the temperature. On the other hand, in the case without the PCM, the HVAC operation results in a much more pronounced temperature variation, which can negatively affect the occupants' thermal comfort. 

In the next section, we detail the formulation of the optimisation problem in HEMS-PCM and briefly explain the computationally efficient method that we proposed to solve the optimisation problem of HEMS-PCM. In particular, we explain how we use the generated training data from the thermal model to solve the HEMS-PCM optimisation problem.

\section{Home energy management in HEMS-PCM} 
\label{sec:Home energy management in HVAC PV-PCM buildings}
In this section, we first present the mathematical formulation of the HEMS-PCM optimisation problem and then briefly describe the methodology that we developed in our earlier work as a method to solve the optimisation problem \cite{rahimpour2021comput}.

\subsection{Optimisation problem formulation}
To solve an optimisation problem with our proposed method of MADP, which is a type of approximate dynamic programming (ADP), we need to formulate the problem as Markov decision processes (MDP).  An MDP comprises a \textit{state-space}, $(s\in \mathcal{S})$, a \textit{decision-space}, $(x\in \mathcal{X})$,  \textit{transition functions} and \textit{cost functions}. 
Let $k=\left\lbrace 1,\ldots,K \right\rbrace$ denote a time-step of half an hour. A state variable, $s_{k}\in \mathcal{S}$, contains the information that is necessary and sufficient to make the decisions and compute costs, rewards and transitions. In this problem, the state variable is the indoor temperature of the building. The decision variable, $x_{k}\in \mathcal{X}$, is an action that results in a transition from one state to another in a sequence over the decision horizon. In this work, action is on/off status of the HVAC system we take in each step. 
For simplicity, we treat the problem as deterministic. This means we assume all information such as weather conditions, customer's electricity demand, and electricity time of use tariff is known and not changing during the whole time-horizon of the problem. Thus the HEMS's objective is
\begin{align}\label{eqn9}
	\underset{\pi}{\text{min}}\ 
	& \mathbb{E}\left\{ \sum^{K}_{k=0} C^{\pi}_{k}({s}_{k},{x}_{k})\right\} \nonumber \\
	\text{s.t.} 
	&\quad \textrm{thermal comfort constraints, and} \nonumber \\ 
	&\quad \textrm{thermal energy balance constraints,}
\end{align}
where $\pi:\mathcal{S}\rightarrow \mathcal{X}$ is a policy, i.e. a sequence of actions taken to move from each state to the next state over the whole time horizon. In this work, a policy is a sequence of on/off status of the HVAC system over a defined time horizon. 

\begin{figure}[t]
	\centering
	\centerline{\includegraphics[width=85mm,keepaspectratio]{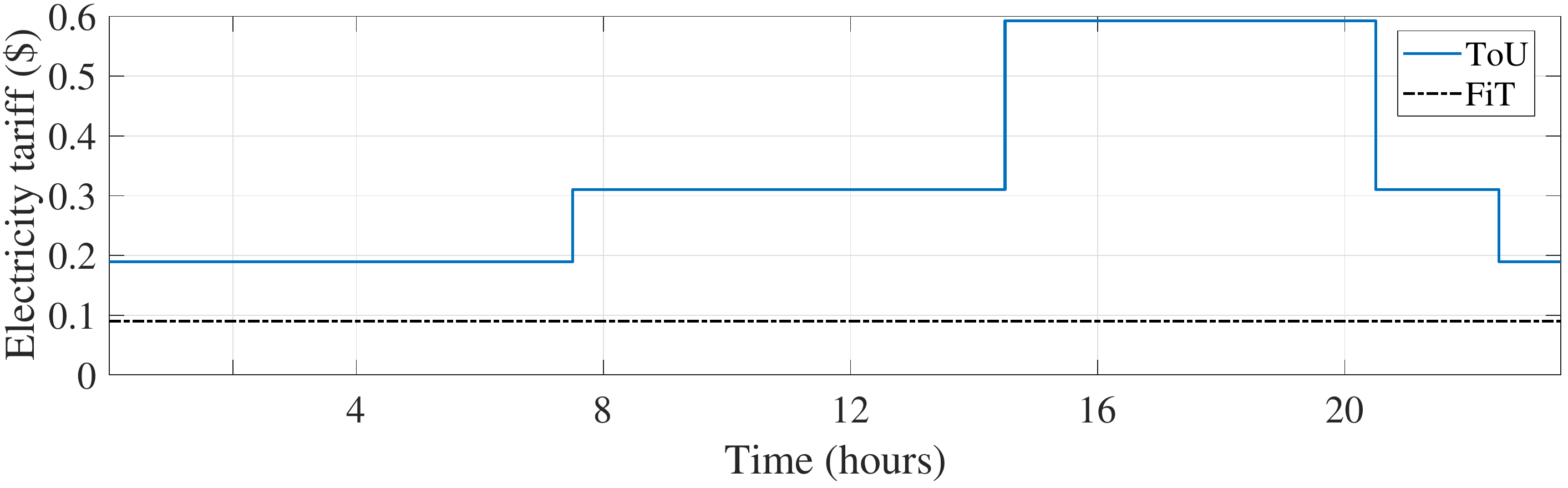}}%
			\caption{The electricity time-of-use tariff and the feed-in-tariffs.}
	\label{fig:tariff2}
\end{figure}

Let ${{s}_{k+1}={s}^{M}\left({s}_{k},{x}_{k}\right)}$ describe the evolution from time step $k$ to the next time step, $k+1$, where ${s}^{M}$ is the underlying mathematical model of the studied system (see \cite{keerthisinghe2018fast} for a detailed HEMS formulation). In this problem, the system model is a system of the ordinary differential equations (ODE) \eqref{eqn7} and \eqref{eqn8}.
Cost function $C_{k}({s}_{k},{x}_{k})$ captures the cost incurred at a given time-step $k$ that accumulates over time. The cost function consists of the cost of \emph{importing} electricity from the grid and the income from \emph{exporting} electricity to the grid:
\begin{equation}\label{eqn10}
	C_{k}({s}_{k},{x}_{k}) = 
	{c}^{\text{ToU}}_{k}{ {p}_{k}^{+} } - {c}^{\text{FiT}}{ {p}_{k}^{-}},
\end{equation}
where ${p}_{k}^{+}$ represents total electricity demand at $k$, and ${p}_{k}^{-}$ is PV generation at $k$. The feed-in tariff $c^{\text{FiT}}$ is assumed to be fixed 0.09 $\SI{}{\$\per\kilo\watt\hour}$, while the electricity tariff ${c}_{\mathrm{g},k}$ is time-of-use, i.e. it changes with time, as shown in Fig.~\ref{fig:tariff2}. Observe that the electricity cost is always higher than the feed-in tariff, which means that the optimal strategy is to use as much as possible the power generated by the PV system; that is, minimising cost is equivalent to \emph{maximising PV self-consumption}.

The cost function only considers the instantaneous cost that results from the decision that is taken at each time step. 
Building on this, dynamic programming solves the optimization problem by computing a \textit{value function} $V^{\pi}({s}_{k})$, which is the expected future discounted cost of following policy $\pi$ starting in state ${s}_{k}$. It is given by:
\begin{align}\label{eqn11}
	{\displaystyle V^{\pi}({s}_{k})
		= \sum_{{s}'\in \mathcal{S}} \mathbb{P}({s}'|{s}_{k},{x}_{k})\left[C({s}_{k},{x}_{k},{s}')+ V^{\pi}({s}')\right]}, 
\end{align} 
where $\mathbb{P}({s}'|{s}_{k},{x}_{k})$ is the transition probability of landing on state ${s}'$ from ${s}_{k}$ if we take action $x_k$. 
However, because the system model, $s^M$ is a deterministic function, we have
\begin{equation}\label{eqn240}
	\mathbb{P}({s}'|{s}_{k},{x}_{k}) = 
	\begin{cases}
		1 & \textrm{if } s^M(s_k,x_k) = s', \\
		0 & \textrm{otherwise,}
	\end{cases}
\end{equation}

The expression in \eqref{eqn11} is a recursive reformulation of the objective function. 
Thus, in general, \textit{Bellman's optimality condition} states that the optimal value function is given by
\begin{align}\label{eqn12} 
	V_{k}^{\pi^{*}}({s}_{k})=\min\limits_{{x}_{k}\in \mathcal{X}_{k}} \left(C_{k}({s}_{k}, {x}_{k}) + \mathbb{E}\left\{ V_{k+1}^{\pi^{*}}({s}')|{s}_{k}\right\}\right),
\end{align}
where $\pi^{*}$ is an optimal policy.
To find $\pi^*$, we need to solve \eqref{eqn12} for each state. 

\textit{Value iteration} is the process of computing \eqref{eqn12} for each state by backward induction; that is, starting at the endpoints of the MDP. 
The optimal policy is extracted from the optimal value function by selecting the minimum value action for each state. 
To describe this in a simple way, in value iteration, the desired state in step $k+1$ is set to the lower value while the undesired states and states that are out of comfort bounds are penalised by assigning higher values. In this work, the occupant's comfort temperature range is considered to be between \SI{20}{\degreeCelsius} to \SI{24}{\degreeCelsius}. 
Then, for all possible states at time $k$, the value iteration algorithm moves backward in time and, in each time step, by solving the subproblem in~\eqref{eqn12}, the minimum value function is computed for different states of each time step. 
In the final step of backward induction, corresponding to the initial starting point, all value function calculations converge to the optimal value function. 
Then, by tracing a minimum value-function path forward for a given time horizon, the optimal policy is found. 

However, despite advancements in computation power, directly applying value iteration (or other exact dynamic programming algorithms) has an excessively high computational burden.  
Although we consider only one state-variable representing the indoor temperature of the building, the running time of the value iteration algorithm for a decision horizon of \SI{24}{hours} with slot length of one hour is very long; in this problem, the running time is almost nine days on a high-performance computer cluster. 
The main reason behind this is that at each time step, the algorithm solves ODEs that govern the thermal behaviour of the model for each action to update the MDP state, and this is repeated until the initial starting state is reached. 
Given this shortcoming, we now briefly describe the method that we proposed in our earlier work to reduce the computational requirements of dynamic programming.

\subsection{Methodology}       
\label{sec:Methodology}
To overcome the high computational burden of dynamic programming, we developed a \textit{multi-time scale approximate dynamic programming} (MADP) algorithm \cite{rahimpour2021comput}. This method incorporates a multi-timescale MDP and an artificial neural network function approximator of the building’s dynamic model, coupled with an underlying state-space approximation. 
In more detail, we used approximation as the foundation of the MADP method. We discretise the state-space of the problem by rounding the output of the ODEs \eqref{eqn7} and \eqref{eqn8}. 
To reduce the runtime of the algorithm further, we use multi-time scale approach \cite{sutton1995td}, in which we solve several smaller MDPs that are connected successively together to form the original MDP, rather than solving the original MDP as one monolithic problem.

To further improve the performance of the algorithm, we address the computational bottleneck of the state transition function that governs the thermal behaviour of the building. 
As an alternative to time-consuming solution of the ODEs, here we propose using an \textit{artificial neural network} (ANN) function approximation of the system of ODEs that maps the HVAC system status (on/off), $x_{k}$, the indoor temperature at $k-1$, $T_{\text{in},k-1}$, the outdoor temperature at $k$, $T_{\text{out},k}$, the outdoor temperature at $k-1$, $T_{\text{out},k-1}$, to the next time-step's indoor temperature.

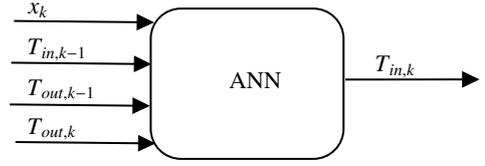
\begin{figure}
	\centering
	\tikzset{every picture/.style={line width=0.75pt}} 
	\begin{tikzpicture}[x=0.75pt,y=0.75pt,yscale=-1,xscale=1,scale=0.70,every node/.style={scale=0.90}]
		
		\draw   (241,122.7) .. controls (241,110.77) and (250.67,101.1) .. (262.6,101.1) -- (356.9,101.1) .. controls (368.83,101.1) and (378.5,110.77) .. (378.5,122.7) -- (378.5,187.5) .. controls (378.5,199.43) and (368.83,209.1) .. (356.9,209.1) -- (262.6,209.1) .. controls (250.67,209.1) and (241,199.43) .. (241,187.5) -- cycle ;
		\draw    (143.5,110) -- (240.5,110.97) ;
		\draw [shift={(243.5,111)}, rotate = 180.57] [fill={rgb, 255:red, 0; green, 0; blue, 0 }  ][line width=0.08]  [draw opacity=0] (12.5,-6.01) -- (0,0) -- (12.5,6.01) -- cycle    ;
		\draw    (141.5,140) -- (238.5,140.97) ;
		\draw [shift={(241.5,141)}, rotate = 180.57] [fill={rgb, 255:red, 0; green, 0; blue, 0 }  ][line width=0.08]  [draw opacity=0] (12.5,-6.01) -- (0,0) -- (12.5,6.01) -- cycle    ;
		\draw    (140.5,170) -- (237.5,170.97) ;
		\draw [shift={(240.5,171)}, rotate = 180.57] [fill={rgb, 255:red, 0; green, 0; blue, 0 }  ][line width=0.08]  [draw opacity=0] (12.5,-6.01) -- (0,0) -- (12.5,6.01) -- cycle    ;
		\draw    (144.5,197) -- (241.5,197.97) ;
		\draw [shift={(244.5,198)}, rotate = 180.57] [fill={rgb, 255:red, 0; green, 0; blue, 0 }  ][line width=0.08]  [draw opacity=0] (12.5,-6.01) -- (0,0) -- (12.5,6.01) -- cycle    ;
		\draw    (379.75,152.1) -- (472.5,152) ;
		\draw [shift={(475.5,152)}, rotate = 539.94] [fill={rgb, 255:red, 0; green, 0; blue, 0 }  ][line width=0.08]  [draw opacity=0] (12.5,-6.01) -- (0,0) -- (12.5,6.01) -- cycle    ;
		
		\draw (295,145) node [anchor=north west][inner sep=0.75pt]   [align=left] {ANN};
		\draw (151,121) node [anchor=north west][inner sep=0.75pt]    {$T_{in,k-1}$};
		\draw (151,150) node [anchor=north west][inner sep=0.75pt]    {$T_{out,k-1}$};
		\draw (151,178) node [anchor=north west][inner sep=0.75pt]    {$T_{out,k} \ $};
		\draw (151,95) node [anchor=north west][inner sep=0.75pt]    {$x_{k}$};
		\draw (399,133) node [anchor=north west][inner sep=0.75pt]    {$T_{in,k}$};
	\end{tikzpicture}
	\vspace{0.2cm}
	\caption{The ANN transition function approximator. The ANN has four inputs, including HVAC system status (on/off), $x_{k}$, the indoor temperature at $k-1$, $T_{\text{in},k-1}$, the outdoor temperature at $k$, $T_{\text{out},k}$, the outdoor temperature at $k-1$, $T_{\text{out},k-1}$, while the output is the indoor temperature at $k$, $T_{\text{in},k}$.}
	\vspace{-0.2cm}
	\label{fig:NN}
\end{figure}

The ANN transition function approximator, illustrated in Fig.~\ref{fig:NN}, can be trained offline and employed as a lookup table within the MADP method. The training dataset is generated using the ODEs of the thermal model of the building presented in Section \ref{sec:Thermal RC lumped model of PCM-buildings}. The results demonstrate that the proposed method performs well with a combined computational speed-up of up to 157,600 times compared to the direct application of dynamic programming.
For the sake of simplicity we assumed that all HEMS inputs (temperature, PV generation, demand and electricity tariff) are deterministic.

\section{Case studies} 
\label{sec:Case studies}
We perform simulations for five Australian capital cities (Sydney, Brisbane, Melbourne, Adelaide and Perth). For each city, we analysed 50 buildings with the same construction (and hence the same thermal performance) but with different demand and solar generation profiles. For Perth, we consider only 10 buildings due to scarcity of data.

\subsection{Demand data}
We assume identical underlying electricity demand for each city that excludes the HVAC demand.
The electricity demand of the HVAC system is the output of either a deadband controller or the HEMS, depending on the scenario.
To capture the variability that exists in real-world customer demand profiles, we generate fifty random demand profiles, one for each house in each city, using the Bayesian non-parametric method developed in our previous work \cite{power2019nonparametric}.
The method first constructs a Markov chain model using the electricity demand from the empirical data (Ausgrid's Smart Grid, Smart City data\footnote{\href{https://data.gov.au/dataset/ds-dga-4e21dea3-9b87-4610-94c7-15a8a77907ef/details}{www.data.gov.au}} in our case). Then, it synthesizes statistically representative demand profiles for an individual house by subsequently sampling the Markov chain model.
The generated demand profiles are such that the aggregated demand profile matches well with the aggregated demand profile of the observed data in the Smart Grid Smart City dataset. The demand profile represents the electricity consumption of a medium residential building in Australia, with an average annual electricity demand of approximately \SI{4.7}{MWh} (around \SI{12.8}{kWh} per day). 

\subsection{Weather data}
The temperature data is from the Australian Bureau of Meteorology. We used outdoor dry bulb temperature data from the neighbouring weather stations (Observatory Hill for Sydney, Brisbane City for Brisbane, Olympic Park for Melbourne, West Terrace for Adelaide and Perth Metro for Perth). The data is for calendar year 2019. The minimum, maximum and average dry bulb outdoor temperatures are given in Table~\ref{table6}. 

\begin{table}[t]
	\footnotesize
	\centering
	\caption{Minimum, maximum and average dry bulb outdoor temperature for calendar year 2019 from the Australian Bureau of Meteorology.}
	\label{table6}
	\begin{tabular}{lccc} 
		\toprule
		\begin{tabular}[c]{@{}l@{}}\textbf{City}\end{tabular} & \begin{tabular}[c]{@{}c@{}}\textbf{Minimum (\SI{}{\degreeCelsius})}\end{tabular} & \begin{tabular}[c]{@{}c@{}}\textbf{Maximum (\SI{}{\degreeCelsius})}\end{tabular} & \begin{tabular}[c]{@{}c@{}}\textbf{Average (\SI{}{\degreeCelsius})}\end{tabular} \\ 
		\midrule
		\textbf{Sydney}    & 6.2 & 39.3 & 18.8 \\
		\textbf{Brisbane}  & 7.6 & 41.0 & 21.6 \\
		\textbf{Melbourne} & 2.4 & 43.2 & 15.7 \\
		\textbf{Adelaide}  & 2.3 & 46.2 & 17.5 \\
		\textbf{Perth}     & 2.1 & 41.8 & 18.6 \\
		\bottomrule
	\end{tabular}
\end{table}

\subsection{PV generation data}
For PV generation, we use data provided by Solar Analytics.\footnote{\href{https://www.solaranalytics.com/}{www.solaranalytics.com}} 
We selected fifty postcodes located in each city and extracted the corresponding solar profiles from the Solar Analytics dataset. Average annual values of PV system generation ${p}_{\mathrm{ave}}^{-}$, is given in Table~\ref{table4}.

\subsection{Output variables}
Variables of interest in our study are (i) electricity cost-saving; 
(ii) PV self-consumption, defined as:
\begin{equation}\label{eqn14}
	SC
	=\frac{\sum^{K}_{k=1}\mathrm{min}({p}_{k}^{+},{p}_{k}^{-})}{\sum^{K}_{k=1}{p}_{k}^{-}}\times 100,
\end{equation}
where ${p}_{k}^{+}$ represents total electricity demand at time $k$, and ${p}_{k}^{-}$ is PV generation at time $k$.

\begin{table*}
    \footnotesize
	\centering
	\caption{Electricity cost-saving and PV self-consumption reduction for PCM melting points of \SI{21}{\degreeCelsius} (MT21) and \SI{23}{\degreeCelsius} (MT23), and PV sizes of \SI{5}{\kilo\watt} and \SI{8}{\kilo\watt}.}
	\label{table7}
	\begin{tabular}{lcclccllllll} 
		\toprule
		& \multicolumn{5}{c}{Electricity cost-saving} & \multicolumn{1}{c}{} & \multicolumn{5}{c}{PV self-consumption reduction} \\ 
		\cmidrule{2-6}\cmidrule{8-12}
		\textbf{City} & \multicolumn{2}{c}{\SI{5}{\kilo\watt}} & & \multicolumn{2}{c}{\SI{8}{\kilo\watt}} & \multicolumn{1}{c}{} & \multicolumn{2}{c}{\SI{5}{\kilo\watt}} & & \multicolumn{2}{c}{\SI{8}{\kilo\watt}} \\ 
		\cmidrule{2-3}\cmidrule{5-6}\cmidrule{8-9}\cmidrule{11-12}
		& \multicolumn{1}{l}{MT21} & \multicolumn{1}{l}{MT23} & & \multicolumn{1}{l}{MT21} & \multicolumn{1}{l}{MT23} & & MT21 & MT23 & & MT21 & MT23 \\ 
		\cmidrule{2-3}\cmidrule{5-6}\cmidrule{8-9}\cmidrule{11-12}
		& $\bar{x}(\$)$ & $\bar{x}(\$)$ & & $\bar{x}(\$)$ & $\bar{x}(\$)$ & & $\bar{x}(\%)$ & $\bar{x}(\%)$ & & $\bar{x}(\%)$ & $\bar{x}(\%)$ \\ 
		\midrule
		\textbf{Sydney} & 174.98 & 113.09 & & 167.64 & 107.81 & & 1.50 & 1.49 & & 1.13 & 1.05 \\
		\textbf{Brisbane} & 100.58 & 98.23 & & 91.72 & 88.73 & & 1.45 & 2.73 & & 0.99 & 1.91 \\
		\textbf{Melbourne} & 306.08 & 180.40 & & 291.90 & 171.91 & & 2.30 & 1.66 & & 1.77 & 1.18 \\
		\textbf{Adelaide} & 253.94 & 168.89 & & 239.40 & 158.00 & & 2.56 & 2.64 & & 1.89 & 1.83 \\
		\textbf{Perth} & 153.89 & 109.35 & & 138.09 & 81.56 & & 2.65 & 3.73 & & 1.92 & 1.84 \\
		\bottomrule
	\end{tabular}
\end{table*}

\subsection{Selection of PCM melting point}
To select the optimal PCM melting point, we investigate its impact on electricity cost-saving and PV self-consumption. To do so, we run yearly simulations for all the sites in all the cities, considering melting points of \SI{21}{\degreeCelsius} (MT21), and \SI{23}{\degreeCelsius} (MT23). The results are summarised in Table~\ref{table7}. In all the cities except Brisbane, the melting point temperature of \SI{21}{\degreeCelsius} results in more than \SI{40}{\%} in electricity cost-saving; for Melbourne, this value as high as \SI{70}{\%}. In Brisbane, MT21 reduces the electricity cost by only \SI{2.4}{\%}. Overall, the difference between in cost-savings between MT21 and MT23 is smaller in cities with warmer winters compared to cities with colder winters. On the other hand, different melting points have a small impact on the reduction in PV self-consumption. The same trend is observed for two different PV capacities (\SI{5}{kW} and \SI{8}{kW}). 
Against this backdrop, we chose MT21 for further analysis. 

\subsection{Simulation scenarios}
The simulation scenarios consider two different HVAC controls, HEMS and deadband control (DB), both with or without PCM; this gives four scenarios, capturing all four combinations:
\begin{enumerate}
\itemsep 0em 
    \item \textbf{DB} (deadband control without PCM)
    \item \textbf{DB-PCM} (deadband control with PCM)
    \item \textbf{HEMS} (HEMS without PCM)
    \item \textbf{HEMS-PCM} (HEMS with PCM)
\end{enumerate}
The setpoint is \SI{21}{\degreeCelsius} and \SI{23}{\degreeCelsius} for the heating and cooling mode, respectively, with a deadband of $\pm\SI{1}{\degreeCelsius}$.
All simulations are run for a whole year with a half-hourly resolution. 

\section{Results and discussion}
\label{sec:Results and discussion}
The results are split into two parts: first, we compare all four scenarios in terms of the HVAC consumption and PV self-consumption, which serves as a baseline for further analysis. Next, we focus specifically on the impact of PCMs on the performance of the HEMS used to control the HVAC system.

\subsection{Scenario comparison: summary statistics}
Table~\ref{table4} summarises the performance of all four scenarios for all five cities, showing PV generation ${p}_{\mathrm{ave}}^{-}$, underlying electricity demand ${d}_{\mathrm{ave}}$ (the same for all sites), HVAC consumption ${d}_{\mathrm{HVAC,ave}}$, and PV self-consumption ${SC}_{\mathrm{ave}}$. The reported values are averages across all sites (fifty for Sydney, Brisbane, Melbourne and Adelaide, and ten for Perth).

As expected, PCMs reduce the HVAC consumption both for HEMS and deadband control. 
However, there is a significant difference between the two control approaches. For deadband control, PCMs reduce the HVAC consumption by \SI{4.2}{\%} on average, with a maximum of \SI{6.7}{\%} in Brisbane and a minimum of \SI{1.9}{\%} in Melbourne.
For HEMS control, on the other hand, PCMs reduce the HVAC consumption by \SI{31.9}{\%} on average, varying between \SI{28.2}{\%} in Brisbane and \SI{34}{\%} in Sydney. 
The reduction in HVAC consumption is even more pronounced when comparing HVAC and deadband control in buildings with PCM. Adding a HEMS reduces the HVAC consumption by \SI{37.2}{\%} on average, with a maximum of \SI{39.3}{\%} in Sydney and a minimum of \SI{32.4}{\%} in Brisbane.

By contrast, PCMs reduce PV self-consumption in all cases except for Brisbane, where PCMs increase PV self-consumption by \SI{0.5}{\%} in the case with the deadband control.
However, the reduction in PV self-consumption is higher when HEMS is used for HVAC control; on average, PCMs reduce PV self-consumption by \SI{2.1}{\%}, whereas the reduction with the deadband control is only \SI{0.5}{\%}.  

\begin{table*}
\footnotesize
\centering
\caption{Scenario comparison for a 5kW PV system showing PV generation (${p}_{\mathrm{ave}}^{-}$), underlying electricity demand (${d}_{\mathrm{ave}}$), HVAC consumption (${d}_{\mathrm{HVAC,ave}}$), and PV self-consumption (${SC}_{\mathrm{ave}}$). The reported values are averages across all sites (fifty for Sydney, Brisbane, Melbourne and Adelaide, and ten for Perth).}
\label{table4}
	\begin{tabular}{lcccccccccc}
		\toprule
		\multirow{2}{*}[-1em]{\textbf{City}} & \multirow{2}{*}[-1em]{$p_{\mathrm{ave}}^{-}$ (kWh)} & \multirow{2}{*}[-1em]{$d_{\mathrm{ave}}$ (kWh)} & \multicolumn{4}{c}{$d_{\mathrm{HVAC,ave}}$ (kWh)} & \multicolumn{4}{c}{$SC_{\mathrm{ave}}$ (kWh)}\\
		\cmidrule(lr){4-7}\cmidrule(lr){8-11}
		& & & \multicolumn{2}{c}{Deadband control} & \multicolumn{2}{c}{HEMS} & \multicolumn{2}{c}{Deadband control} & \multicolumn{2}{c}{HEMS} \\
		\cmidrule(lr){4-7}\cmidrule(lr){8-11}
		& & & NO PCM & PCM &  NO PCM & PCM &  NO PCM & PCM &  NO PCM & PCM \\
		\midrule
		\textbf{Sydney}    & 7916.2 & 4685.3 & 2031.8 & 1952.1 & 1794.5 & 1184.8 & 2284.6 & 2256.9 & 2218.1 & 2099.4 \\
		\textbf{Brisbane}  & 9192.5 & 4685.3 & 1650.9 & 1540.3 & 1449.0 & 1040.7 & 2506.8 & 2557.3 & 2448.9 & 2315.6 \\
		\textbf{Melbourne} & 7332.4 & 4685.3 & 3572.3 & 3503.9 & 3187.3 & 2132.5 & 2361.8 & 2328.8 & 2344.9 & 2176.3 \\
		\textbf{Adelaide}  & 8294.5 & 4685.3 & 3455.3 & 3304.5 & 3053.2 & 2104.8 & 2656.7 & 2581.3 & 2621.1 & 2408.7 \\
		\textbf{Perth}     & 9506.4 & 4685.3 & 2032.2 & 1951.9 & 1799.8 & 1200.1 & 2643.7 & 2627.6 & 2623.8 & 2371.9 \\
		\bottomrule
	\end{tabular}
\end{table*}

\begin{figure*}
	{\centering%
		\begin{tabular}{@{}ll@{}}
			\includegraphics[width=90mm,keepaspectratio]{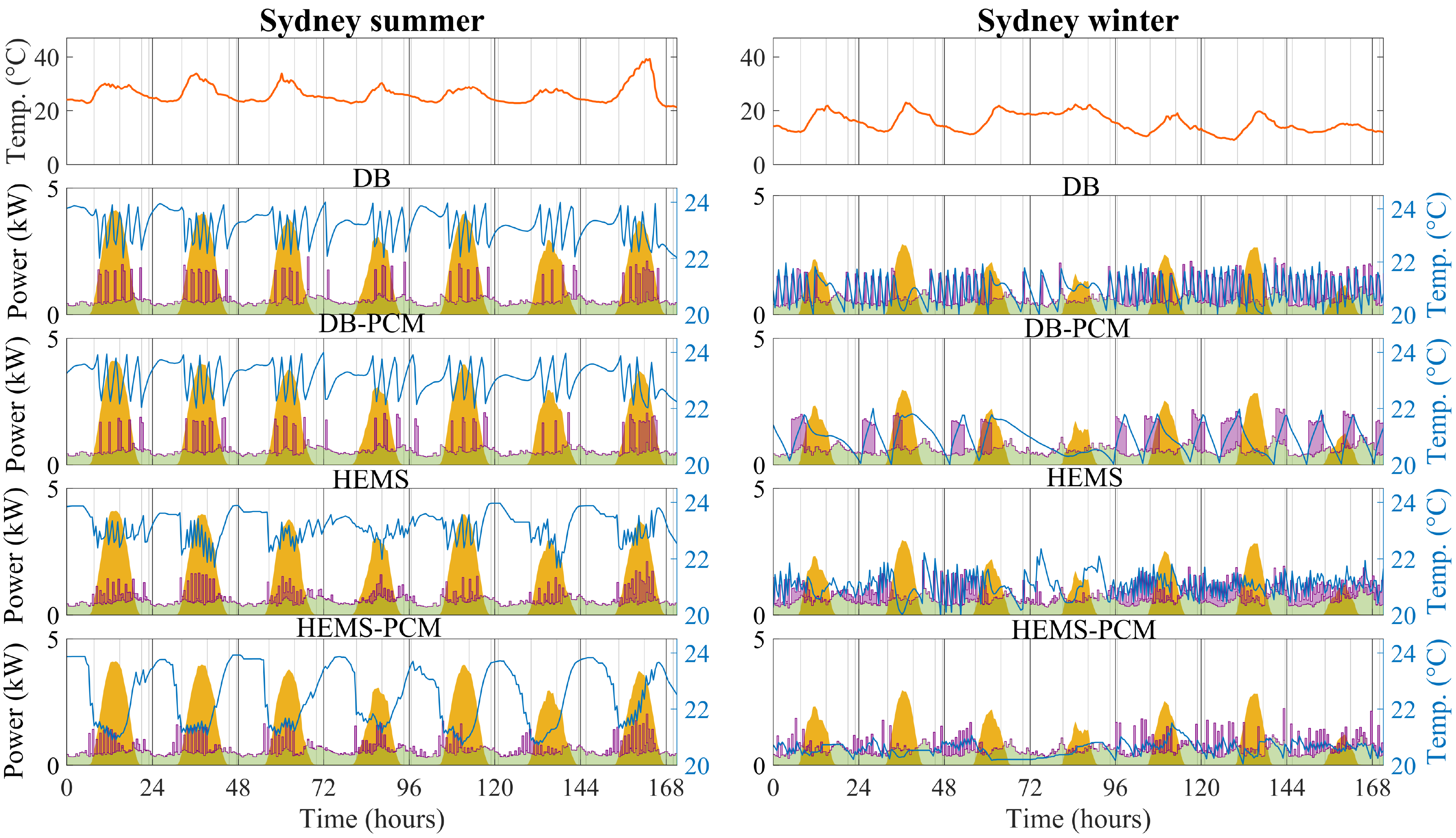}&
			\includegraphics[width=90mm,keepaspectratio]{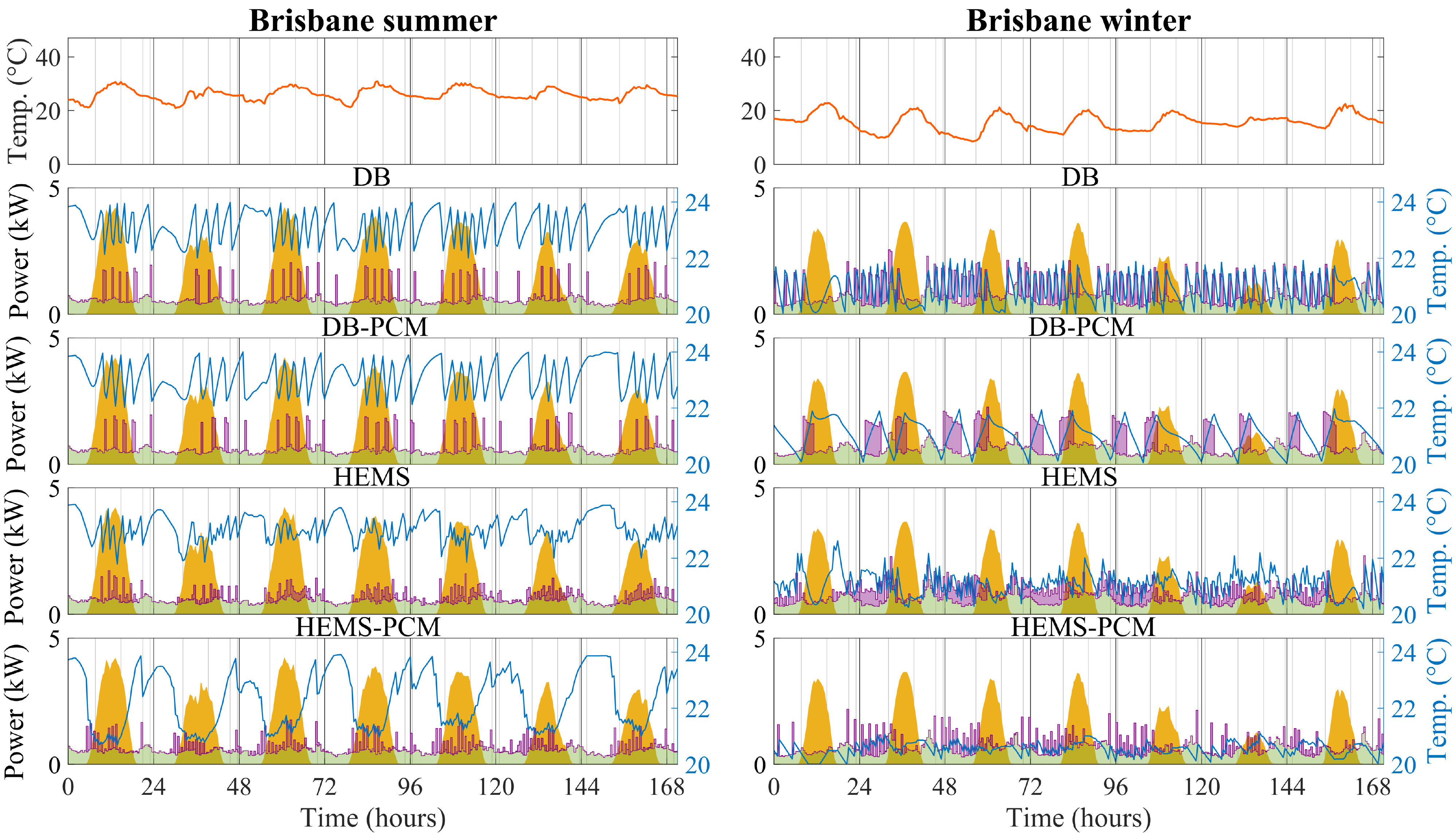}
		\end{tabular}\par}
	{\centering%
		\begin{tabular}{@{}ll@{}}
			\includegraphics[width=90mm,keepaspectratio]{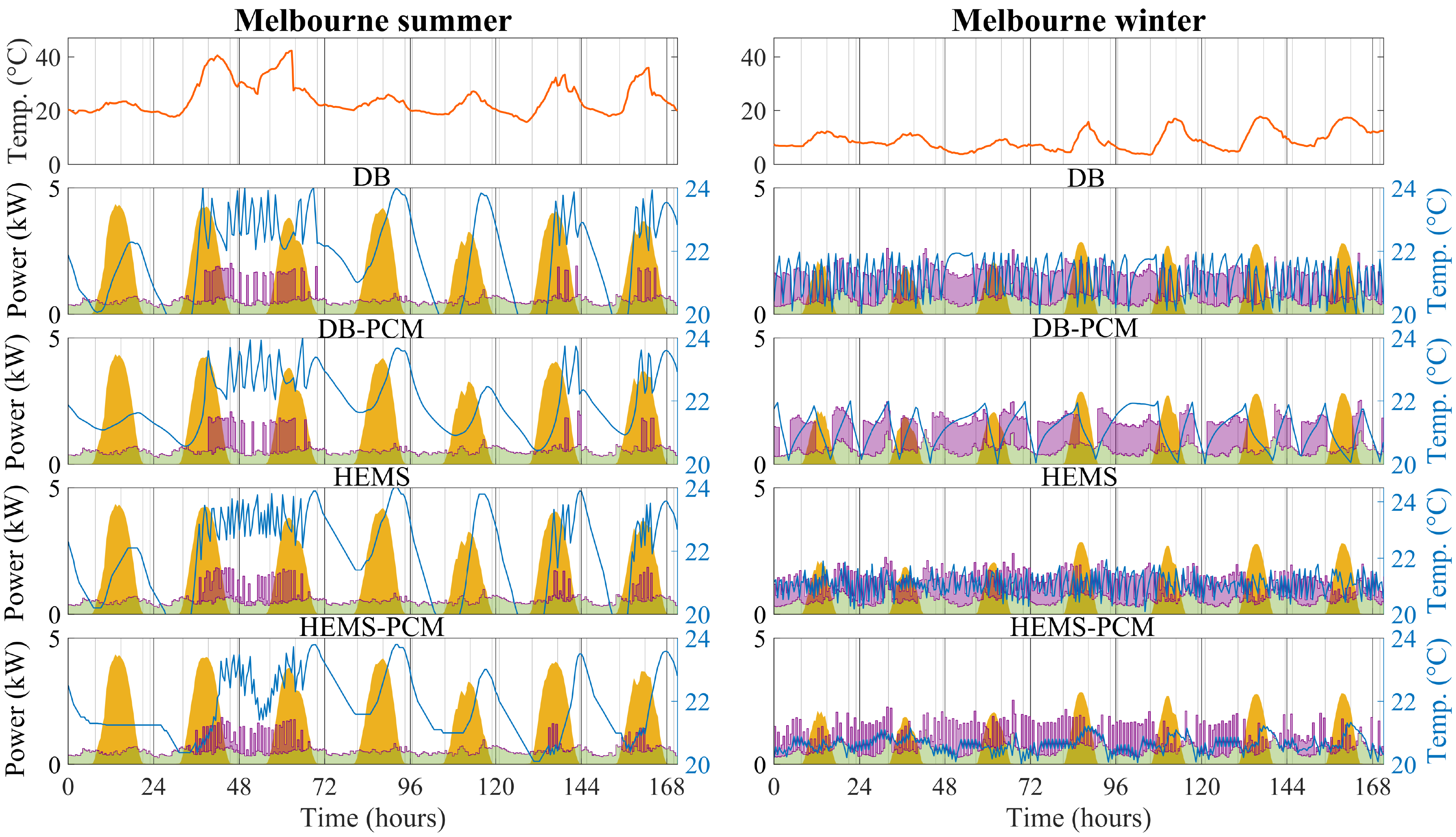}&
			\includegraphics[width=90mm,keepaspectratio]{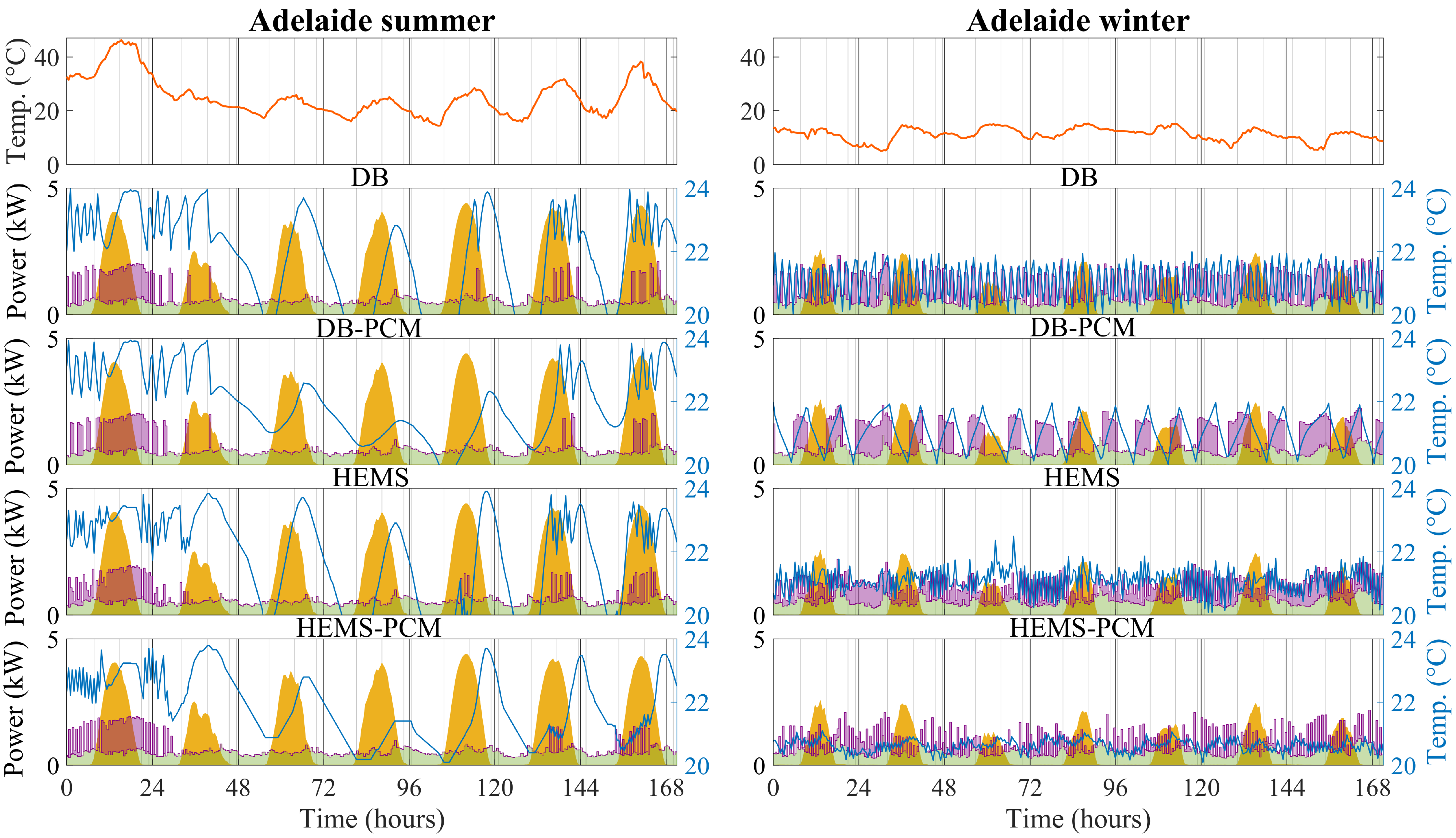}
		\end{tabular}\par}
	{
		\begin{tabular}{@{}lc@{}}
			\includegraphics[width=90mm,keepaspectratio]{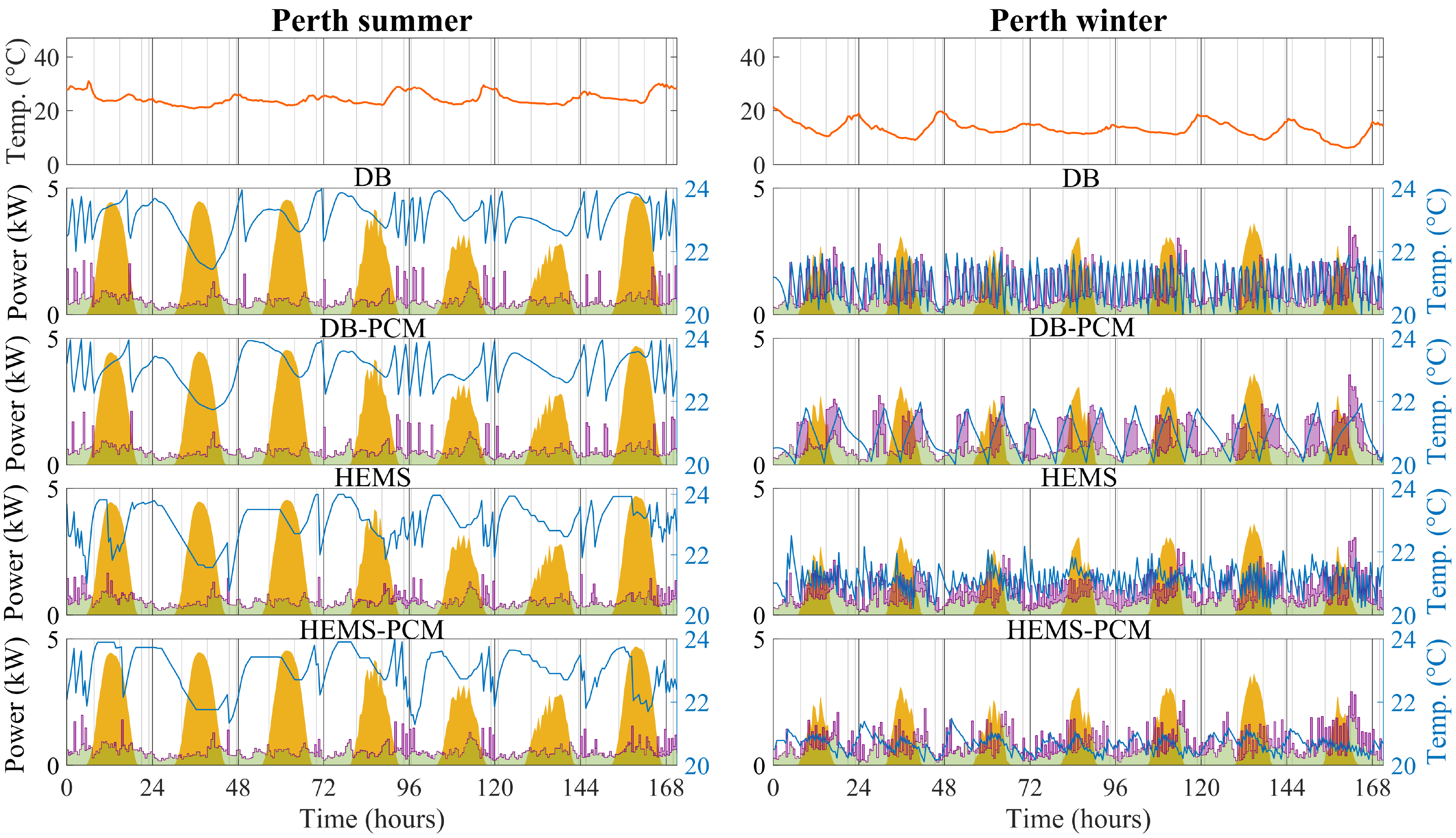}&
			\hspace{1.0em}\centering\includegraphics[width=30mm,keepaspectratio]{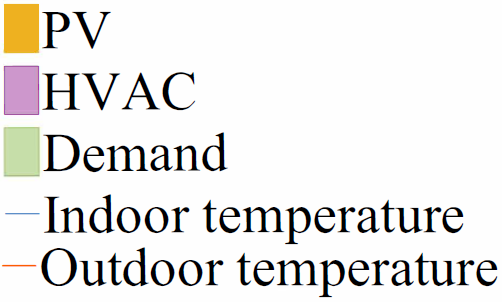}
		\end{tabular}
		\caption{Scenario comparison for Sydney, Brisbane, Melbourne, Adelaide, and Perth for a typical summer week (left), and a typical winter week (right). From top to bottom: outdoor temperature, indoor temperature, PV generation for a \SI{5}{\kilo\watt} system, electricity demand, and HVAC consumption (average values across all the sites in a city). The minor grid on the x-axis indicates the times of the changes in the time-of-use tariff, as shown in Fig.~\ref{fig:tariff2} (22:30-7:30 off-peak, 7:30-14:30 and 20:30-22:30 shoulder, and 14:30-20:30 peak).}
		\label{fig:weekAll}\par}
\end{figure*}

\begin{table*}[t]
    \footnotesize
	\centering
	\caption{Impact of PCMs on HEMS performance: the baseline scenario is a building with a HEMS but no PCM. The reported values are averages across all sites (fifty for Sydney, Brisbane, Melbourne and Adelaide, and ten for Perth). The comparison is done for two performance metrics: electricity cost-saving and PV self-consumption reduction. The comparison is done for two PV system sizes, \SI{5}{\kilo\watt} and \SI{8}{\kilo\watt}.}
\label{table5}
	\begin{tabular}{lcccccccccccccc}
		\toprule
		\multirow{2}{*}[-1em]{\textbf{City}} & \multicolumn{8}{c}{Electricity cost-saving} & \multicolumn{6}{c}{PV self-consumption reduction}\\
		\addlinespace[0pt]
		\cmidrule(lr){2-9} \cmidrule(lr){10-15} \\
		\addlinespace[-12pt]
		{} & \multicolumn{4}{c}{\SI{5}{\kilo\watt}} & \multicolumn{4}{c}{\SI{8}{\kilo\watt}} & \multicolumn{3}{c}{\SI{5}{\kilo\watt}} & \multicolumn{3}{c}{\SI{8}{\kilo\watt}} \\			
		\cmidrule(lr){2-5} \cmidrule(lr){6-9}  \cmidrule(lr){10-12} \cmidrule(lr){13-15}\\
		\addlinespace[-12pt]	
		{} & $\bar{x}(\%)$ & $\bar{x}(\$)$ & $\mathrm{SE}_{\bar{x}}(\$)$ & $\sigma_{x}(\$)$ & $\bar{x}(\%)$ & $\bar{x}(\$)$ & $\mathrm{SE}_{\bar{x}}(\$)$ & $\sigma_{x}(\$)$ & $\bar{x}(\%)$ & $\mathrm{SE}_{\bar{x}}(\%)$ & $\sigma_{x}(\%)$ & $\bar{x}(\%)$ & $\mathrm{SE}_{\bar{x}}(\%)$ & $\sigma_{x}(\%)$ \\
		\midrule
		\textbf{Sydney}    & 13.35 & 174.98 & 0.90 & 6.34  & 17.56 & 167.64 & 1.01 & 7.15  & 1.50 & 0.02 & 0.16 & 1.13 & 0.01 & 0.11\\
		\textbf{Brisbane}  & 10.60  & 100.57  & 0.83 & 5.90  & 16.96 & 94.46   & 0.71 & 5.05 & 1.45 & 0.03 & 0.18 & 1.05 & 0.02 & 0.15\\
		\textbf{Melbourne} & 16.18 & 306.10 & 1.76 & 12.43  & 18.51 & 291.90 & 2.10 & 14.89  & 2.30 & 0.03 & 0.18 & 1.77 & 0.03 & 0.18\\
		\textbf{Adelaide}  & 18.96 & 253.97 & 1.33 & 9.38  & 26.72 & 239.41 & 1.33 & 9.38  & 2.55 & 0.03 & 0.19 & 1.89 & 0.02 & 0.14\\
		\textbf{Perth}     & 16.62 & 153.88 & 9.66 & 30.55 & 27.57 & 138.08 & 10.24 & 32.40 & 2.65 & 0.19 & 0.59 & 1.92 & 0.14 & 0.43\\
		\bottomrule
	\end{tabular}
\end{table*}

\subsection{Scenario comparison: typical summer and winter weeks}
Fig.~\ref{fig:weekAll} illustrates the performance of each scenario for a typical summer and winter week in each city. Each plot shows outdoor and indoor temperatures, PV generation, electricity demand and HVAC consumption (average values across all the sites in a city). The minor grid on the x-axis indicates the times of the changes in the time-of-use tariff, as shown in Fig.~\ref{fig:tariff2} (22:30-7:30 off-peak, 7:30-14:30 and 20:30-22:30 shoulder, and 14:30-20:30 peak) to highlight the load-shifting potential of a HEMS.

A few trends are notable.
First, Fig.~\ref{fig:weekAll} clearly shows the reduced HVAC demand due to the PCM, which is more pronounced in the HEMS cases. 
Second, the PCM smooths out temperature fluctuations, which is due to the increased thermal inertia. This is most obvious in the DB cases; the PCM’s thermal inertia results in a much less frequent on/off toggling of the HVAC system, which results in a smoother temperature profile. 
Third, controlling the HVAC system by the HEMS shifts the HVAC demand from peak hours to off-peak and shoulder hours to reduce the electricity cost. That is quite pronounced, for example, in winter in Brisbane. Comparing DB and HEMS cases shows that the demand in the peak period (14:30-20:30) in the HEMS case is significantly reduced and shifted to the shoulder period (7:30-14:30).
Finally, the high thermal inertia due to the PCM enables preheating and precooling, which results in an even more pronounced shift in the HVAC demand. That is clearly visible, for example, in summer in Sydney (compare HEMS with HEMS-PCM). Observe how the HVAC demand in the HEMS-PCM case is shifted from the peak period (14:30-20:30) to the shoulder period (7:30-14:30).

Fig.~\ref{fig:weekAll} also illustrates how the choice of the PCM melting temperature affects the optimal HEMS performance. Consider, for example, the summer week in Sydney. Observe how the HVAC demand in the HEMS-PCM case is shifted to earlier in the day compared to the HEMS case. That behaviour appears sub-optimal because running the HVAC in the middle of the day to use the free PV generation would cost less. This seemingly suboptimal behaviour can be explained by considering the thermal properties of the PCM. When the PCM is fully melted, it acts as an additional insulation layer that traps the heat. Because of that, the HVAC has to cool the building down to the PCM melting point (\SI{21}{\degreeCelsius} in our case) to prevent the indoor temperature from rising above the upper limit of the comfort range (\SI{24}{\degreeCelsius}). That goes to show that a PCM with a higher melting point would perform better in summer. However, because we consider the cost for the whole year, the benefit of a PCM with a lower melting point in winter outweighs the suboptimal performance in summer.

 \subsection{Impact of PCMs on HEMS performance}
Table~\ref{table5} summarises the analysis of the impact of PCMs on the performance of a HEMS using two performance metrics: electricity cost-saving and the reduction in PV self-consumption for two PV system sizes (\SI{5}{kW} and \SI{8}{kW}). 
The table shows for each city: the average value across all sites $\bar{x}$, standard error of the mean value $\mathrm{SE}_{\bar{x}}$ and standard deviation $\sigma_{x}$. For electricity cost-saving the values are given in \$ and also in \%, while for the PV self-consumption reduction, all the values are given in \%.

\begin{figure}[t]
	\centering
	\centerline{\includegraphics[width=90mm,keepaspectratio]{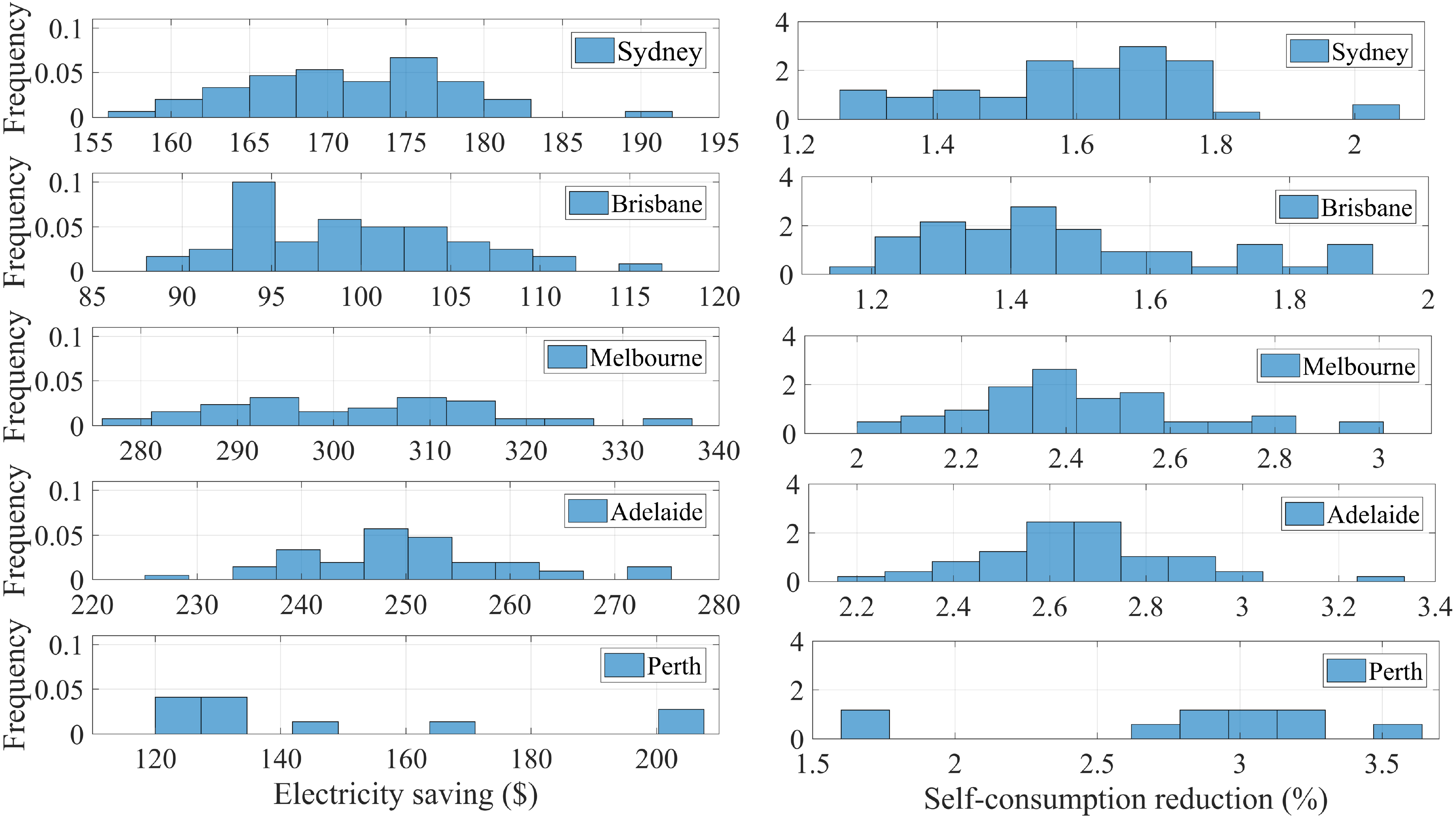}}%
			\caption{Comparison of HEMS with PCM vs HEMS without PCM: histogram of cost-saving (left), and  reduction in PV self-consumption (right) for a PV system size of 5 kW.}
	\label{fig:figcost5}
\end{figure}

\begin{figure}[t]
	\centering
	\centerline{\includegraphics[width=90mm,keepaspectratio]{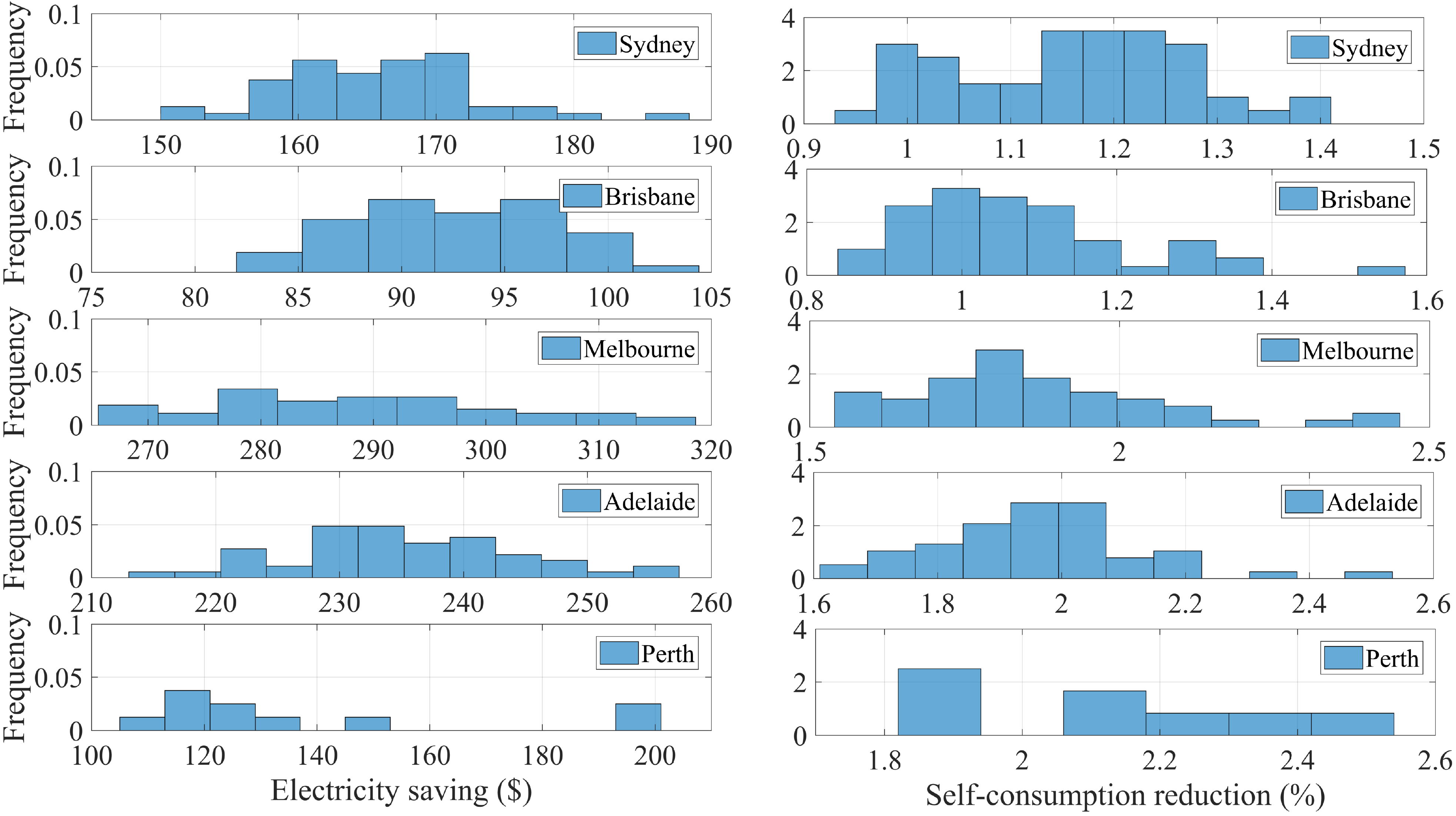}}%
			\caption{Comparison of HEMS with PCM vs HEMS without PCM: histogram of cost-saving (left), and  reduction in PV self-consumption (right) for a PV system size of 8 kW.}
	\label{fig:figcost8}
\end{figure}

The key observation is that PCMs reduce the electricity cost by between \SI{10.6}{\%} in Brisbane and \SI{19}{\%} in Adelaide. Increasing the PV size from \SI{5}{\kilo\watt} to \SI{8}{\kilo\watt} reduces the electricity cost even further, by between \SI{17}{\%} in Brisbane and \SI{27.6}{\%} in Perth. By contrast, PCMs reduce self-consumption by between \SI{1.5}{\%} in Brisbane and \SI{2.7}{\%} in Perth. Increasing the PV size to \SI{8}{\kilo\watt} increases PV self-consumption somewhat, but compared to the base case, PV self-consumption is still reduced. The results are summarised in Figs.~\ref{fig:figcost5} and \ref{fig:figcost8} for a \SI{5}{\kilo\watt} and \SI{8}{\kilo\watt} PV system, respectively.

\section{Conclusion} 
\label{sec:Conclusion}

The paper has explored the potential of PCMs in building insulation to serve as a \emph{solar sponge} to increase self-consumption of rooftop solar PV when a home energy management system optimises the operation of the HVAC system. 
To address the non-convexity of the optimisation problem used for HVAC scheduling, we used our previously developed multi-time scale approximate dynamic programming. This method is computationally efficient and can deal with the nonlinear characteristics of the PCM.
Analysis was conducted for five Australian capital cities, using electricity cost reductions and PV self-consumption as performance metrics.

The hypothesis was that the thermal mass of the PCM would allow preheating and precooling of the building by shifting the HVAC operation to the middle of the day, thus \emph{increasing} PV self-consumption. However, the results show that adding PCMs to the building envelope actually \emph{reduces} it. While the HEMS does shift the operation of the HVAC system to midday, this effect is overshadowed by the overall reduction in HVAC operation. Adding a PCM to a building with an HVAC system controlled by a HEMS namely reduces the HVAC consumption by around 30\% in all five cities. While that results in a significant cost-saving, it reduces PV self-consumption by around 1\% to 3\%. The  sensitivity analysis showed that increasing the size of the PV system from \SI{5}{\kilo\watt} to \SI{8}{\kilo\watt} increased the cost-saving, which was expected but had only a limited impact on the PV self-consumption. The PV self-consumption increased by less than one  percentage point in all cities but remained lower compared to the base case (HEMS without PCM).

The analysis was performed for fifty residential homes in each city (ten for Perth due to paucity of data) using real-life PV generation and electricity demand data, so the results are statistically representative. While we use a simplified building model, we believe that the results are broadly indicative of the role PCMs can play in home energy management.

In future work, we will improve the accuracy of the thermal model by considering more realistic multi-zone buildings, external heat gains, and householders' behaviour, including the associated heat gains. We will also consider the stochasticity of the input data; we have demonstrated in our previous work \cite{Rahimpour2020EPSR} that this can be done using reinforcement learning. 
Our ultimate goal is to embed the HEMS algorithm in a smart meter or another device running the building controller. Reinforcement learning appears to be a promising approach given its ability to learn an optimal control policy without an explicit building model, which will enable plug-and-play operation with little installation overhead.

\section{Acknowledgement} \label{Sec:Acknowledgement}
The authors acknowledge the Sydney Informatics Hub and the use of the University of Sydney's high performance computing cluster, Artemis. 
We would also like to thank Solar Analytics for providing PV generation data and Ausgrid for providing smart meter data.

\bibliographystyle{elsarticle-num}
\bibliography{cas-refs}

\end{document}